%% file: arxiv.tex
\documentclass[10pt,twocolumn,letterpaper]{article}
\usepackage[table,xcdraw]{xcolor}

\usepackage[pagenumbers]{cvpr} %
\usepackage[accsupp]{axessibility}  %

\usepackage{graphicx}
\usepackage{amsmath}
\usepackage{amssymb}
\usepackage{booktabs}
\usepackage{xcolor}
\usepackage{multirow}
\usepackage[normalem]{ulem}
\usepackage[symbol]{footmisc}

\usepackage[super]{nth}

\usepackage[pagebackref,breaklinks,colorlinks]{hyperref}

\usepackage[capitalize]{cleveref}
\crefname{section}{Sec.}{Secs.}
\Crefname{section}{Section}{Sections}
\Crefname{table}{Table}{Tables}
\crefname{table}{Tab.}{Tabs.}

\def\cvprPaperID{1845} %
\def\confName{CVPR}
\def\confYear{2023}

\begin{document}

\title{Patch-based 3D Natural Scene Generation from a Single Example}

\author{Weiyu Li$^{1}$\footnotemark[1]~\footnotemark[3] \qquad Xuelin Chen$^{2}$\footnotemark[1]~\footnotemark[2] \qquad Jue Wang$^{2}$ \qquad Baoquan Chen$^{3}$
\vspace{6pt}\\
$^{1}$ Shandong University \qquad $^{2}$ Tencent AI Lab \qquad $^{3}$ Peking University
}

\twocolumn[{%
\renewcommand\twocolumn[1][]{#1}%
\maketitle
\begin{center} 
    \centering 
    \includegraphics[width=1.\textwidth]{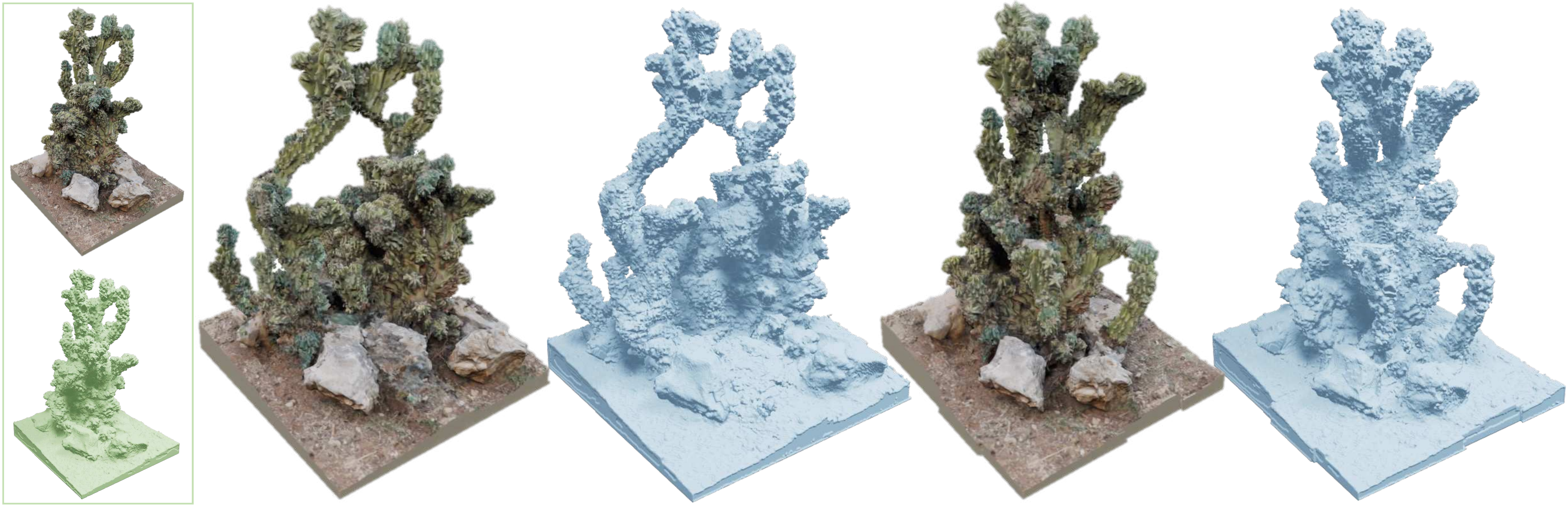} 
    \captionof{figure}{
    From a single example (inset),
    our method can generate diverse samples with highly realistic geometric structure and visual appearance.
    None of existing 3D generative models can achieve these results.
    More generated samples that are intricate and intriguing, ranging from single-object scenes (e.g., plants) to mid-scale ones, and large terrains, 
    can be found in Section~\ref{sec:exp} and the supplementary.
    } 
    \label{fig:teaser}
\end{center}%
}]

\renewcommand{\thefootnote}{\fnsymbol{footnote}} 
\footnotetext[1]{Joint first authors}
\footnotetext[2]{Corresponding author}
\footnotetext[3]{Work done during an internship at Tencent AI Lab.}

\input{0-abs}

\input{1-intro}

\input{2-related}
\input{3-method}

\input{4-exp}

\input{5-application}

\input{6-conclusion}

\textbf{Acknowledgements.} 
This work was supported in part by National Key R\&D Program of China 2022ZD0160801.

{\small
\bibliographystyle{ieee_fullname}
\bibliography{ref}
}

\clearpage
\setcounter{section}{0}
\renewcommand\thesection{\Alph{section}}

\twocolumn[{%
\begin{center} 
    \centering 
    \section*{
    Supplementary Material: \\
        Patch-based 3D Natural Scene Generation from a Single Example
    }
    \includegraphics[width=1.\textwidth]{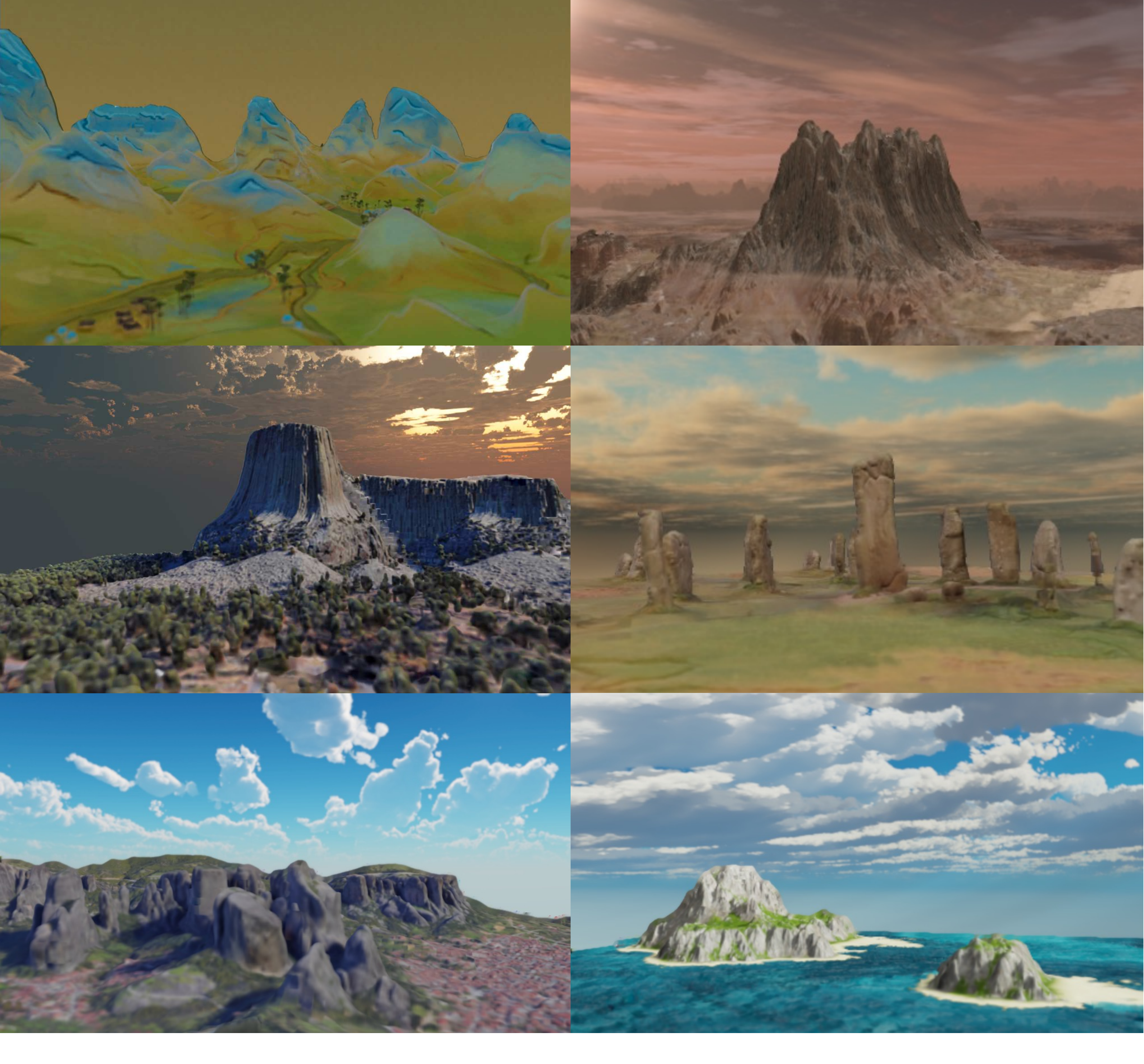} 
    \captionof{figure}{
    Artistic imagery created with various 3D scenes generated by our method (background sky post-added).
    The original exemplar scenes for generating these results are: 
    (from left to right and top to bottom): 
    The Vast Land~\cite{vastland}, 
    Heal Mountain~\cite{healmountain}, 
    Devil's Tower~\href{https://www.google.com/help/terms_maps/}{\textcopyright2022 Google}, 
    Callanish~\cite{callanish}, 
    Meteora~\href{https://www.google.com/help/terms_maps/}{\textcopyright2022 Google}.
    Green Island~\cite{greenisland},
    } 
    \label{fig:supplementary_teaser}
\end{center}%
}]

\input{7-supplementary}

\begin{figure*}[t!]
    \centering
    \includegraphics[width=\hsize]{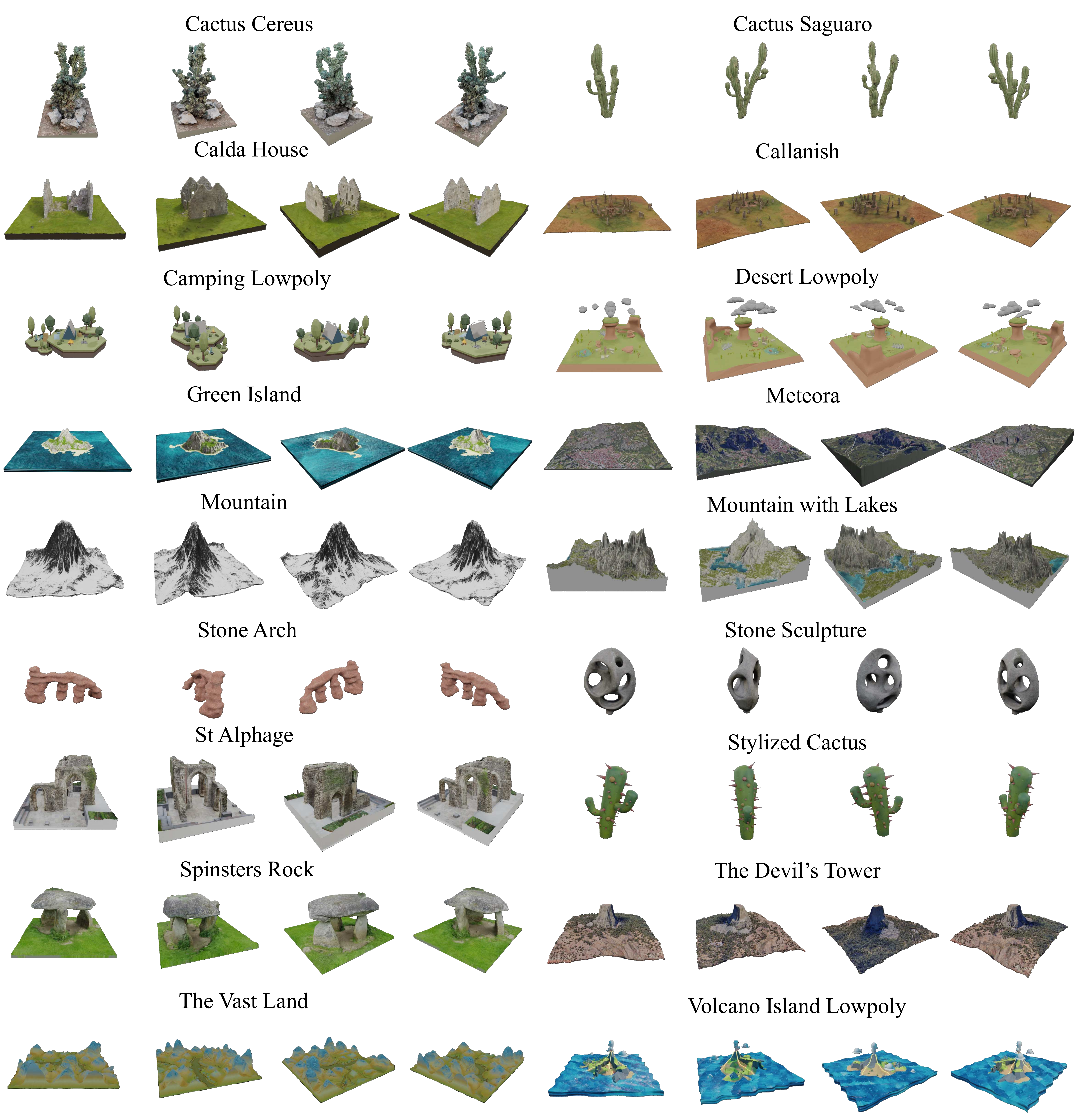}
    \caption{
    Visualization of the exemplars used in the \textbf{main paper}. More scenes can be found in the project page and video.
    }
    \label{fig:supplementary_exemplar}
\end{figure*}
\begin{figure*}[t!]
    \centering
    \includegraphics[width=\hsize]{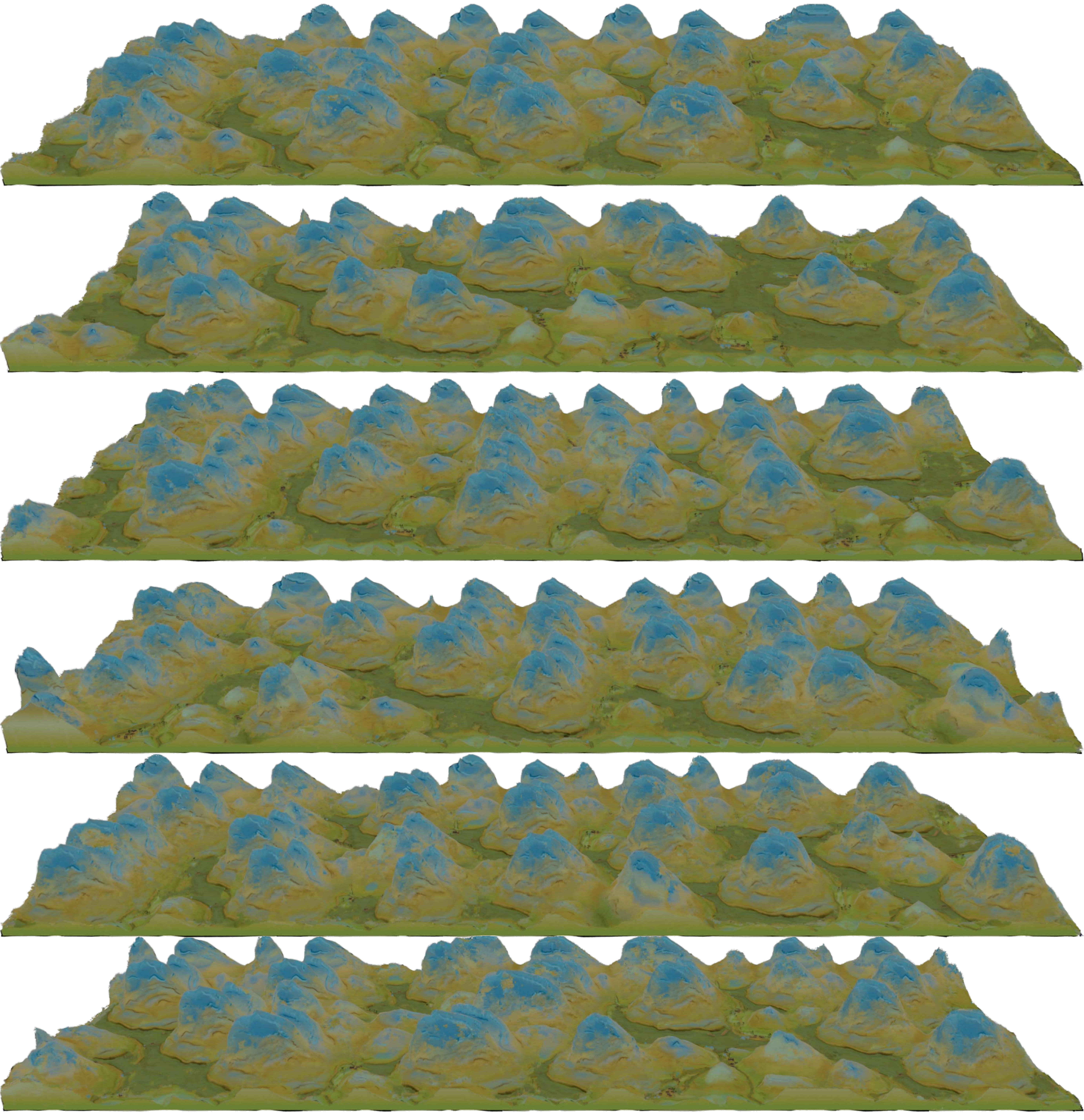}
    \caption{
        Diverse \emph{"A Thousand Li of Rivers and Mountains"}~\cite{wangximeng} generated from The Vast Land~\cite{vastland} by our method.
      Specification:
      $\exemplar_N$ - $288\times288\times112$,
      $\exemplar^{high}$ - $512\times512\times200$,
      $\synthesis_N$ - $747\times288\times112$,
      $\exemplar^{high}(\synthesis_N)$ - $1328\times512\times200$,
      final rendering resolution - $4096\times1024$.
    }
    \label{fig:supplementary_results_1}
\end{figure*}
\begin{figure*}[t!]
    \centering
    \includegraphics[width=\hsize]{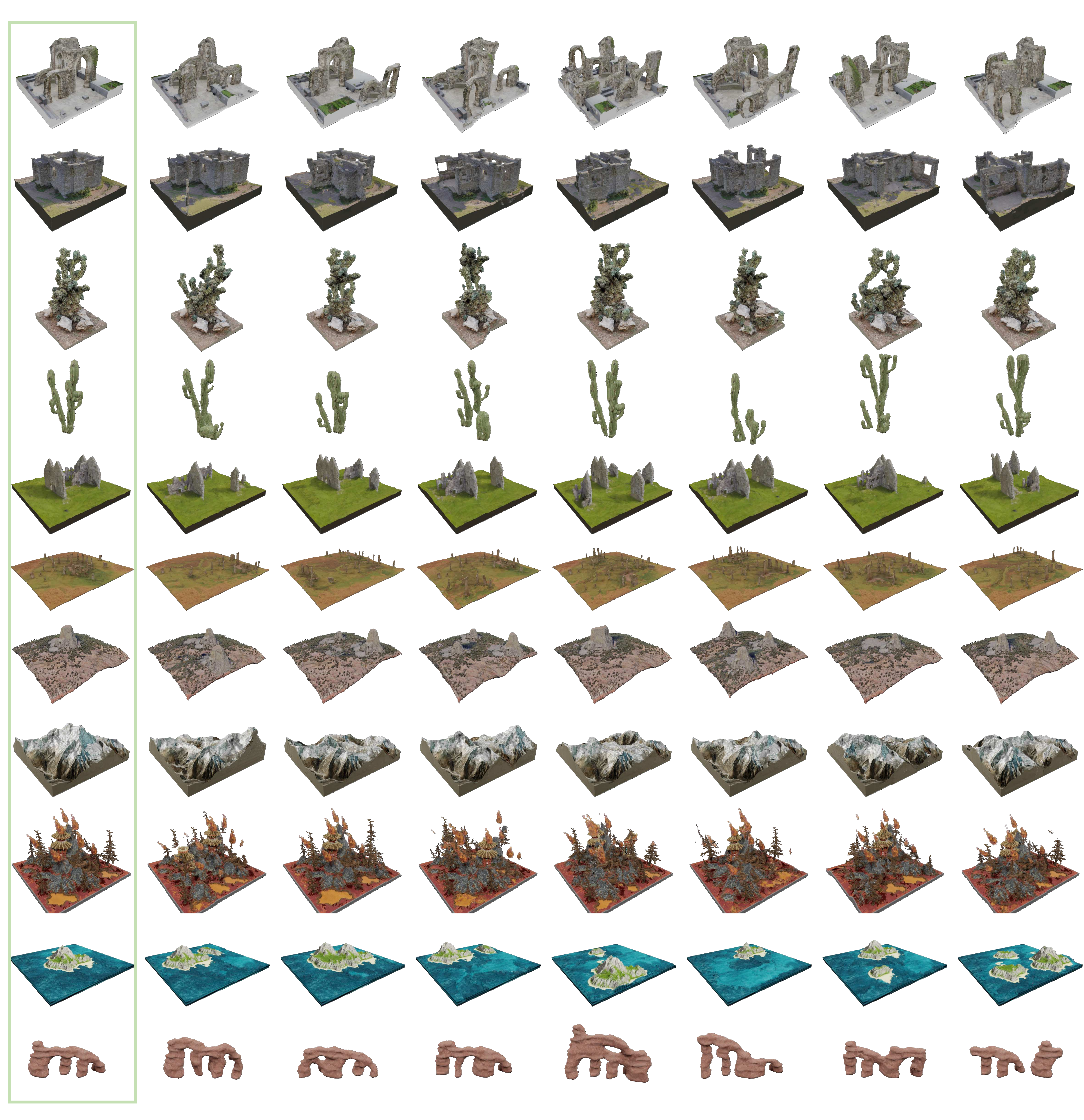}
    \caption{
      Diverse samples generated by our method.
      The input is shown in the green box on the left, followed by 7 generated novel scenes.
    }
    \label{fig:supplementary_results_0}
\end{figure*}
\begin{figure*}[t!]
    \centering
    \includegraphics[width=\hsize]{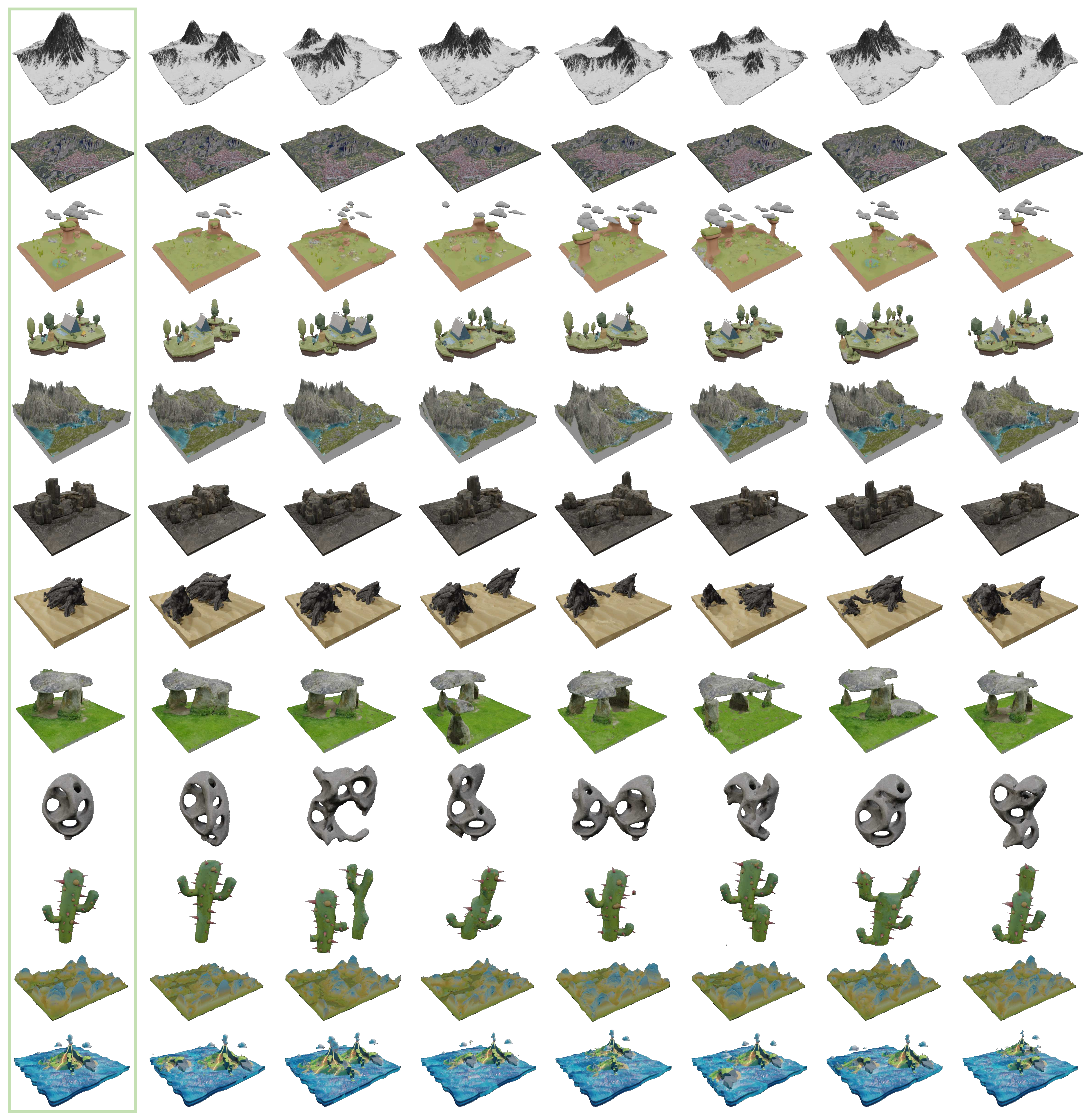}
    \caption{
      Diverse samples generated by our method.
      The input is shown in the green box on the lef,
      followed by 7 generated novel scenes.
    }
    \label{fig:supplementary_results_1}
\end{figure*}

\end{document}

%% file: 0-abs.tex
\begin{abstract}
We target a 3D generative model for general natural scenes
that are typically unique and intricate.
Lacking the necessary volumes of training data,
along with the difficulties of having ad hoc designs in presence of varying scene characteristics,
renders existing setups intractable.
Inspired by classical patch-based image models, 
we advocate for synthesizing 3D scenes at the patch level,
given a single example.
At the core of this work lies important algorithmic designs w.r.t the scene representation 
and generative patch nearest-neighbor module,
that address unique challenges arising from lifting classical 2D patch-based framework to 3D generation.
These design choices, on a collective level, contribute to a robust, effective, and efficient model that can generate high-quality general natural scenes with both realistic geometric structure and visual appearance, in large quantities and varieties,
as demonstrated upon a variety of exemplar scenes.\
Data and code can be found at \href{http://wyysf-98.github.io/Sin3DGen}{http://wyysf-98.github.io/Sin3DGen}.
\end{abstract}

%% file: 1-intro.tex
\section{Introduction}
\label{sec:intro}

3D scene generation generally carries the generation of both realistic geometric structure and visual appearance.
A wide assortment of scenes on earth,
or digital ones across the internet,
exhibiting artistic characteristics and ample variations over geometry and appearance, can be easily listed.
Being able to populate these intriguing scenes in the virtual universe has been a long pursuit in the community.

Research has taken several routes,
among which a prevalent one is learning to extract common patterns of the geometry \emph{or} appearance from homogeneous scene samples,
such as
indoor scenes~\cite{makeithome, wang2018deep, li2019grains, zhang2020deepindoor, wu2019datainterior, fu2017adaptiveindoor, ma2018language, zhu2017indoorcolorization, jain2012materialmemex}, 
terrains~\cite{terrainreview, terraincnn, kapp2020data, hao2021gancraft},
urban scenes~\cite{feng2016crowd, nishida2016interactiveurban, kelly2018frankengan},
etc.
Another line learns to generate single objects~\cite{scores, gao2021tm, chaudhuri2011probabilistic, li2017grass, li2021sp, ngp, park2018photoshape, jain2012materialmemex}.
A dominant trend in recent has emerged that learns 3D generative models to jointly synthesize 3D structures and appearances via differentiable rendering~\cite{graf, giraffe, pi-gan, unconstrained, triplane, stylenerf}.
Nevertheless, 
all these learning setups are limited in their ability to generalize in terms of varied scene types.
While a more promising direction is the exemplar-based one, 
where one or a few exemplars featuring the scene of interest are provided,
algorithm designs tailored for certain scene types in existing methods~\cite{zhou2007terrain, merrell2008continuous, merrell2007example, merrell2010model}
again draw clear boundaries of scene characteristics they can handle.

This work seeks to generate \emph{general natural} scenes,
wherein 
the geometry and appearance of constituents
are often tightly entangled and contribute jointly to unique features.
This uniqueness hinders one from collecting sufficient homogeneous samples for learning common features,
directing us to the exemplar-based paradigm.
On the other hand,
varying characteristics across different exemplar scenes restrain us from having ad hoc designs for a certain scene type (e.g., terrains).
Hence,
we resort to classical patch-based algorithms, which date long before the deep learning era and prevail in several image generation tasks even today~\cite{drop, sketchpatch, vgpnn}.
Specifically,
given an input 3D scene,
we synthesize novel scenes at the patch level 
and particularly adopt the multi-scale generative patch-based framework introduced in~\cite{drop}, 
where the core is a \emph{Generative Patch Nearest-Neighbor} module
that maximizes the bidirectional visual summary\cite{simakov2008summarizing}
between the input and output. 
Nevertheless,
key design questions yet remain in the \emph{3D} generation:
What representation to work with?
And how to synthesize \emph{effectively} and \emph{efficiently}? 

In this work,
we exploit a grid-based radiance field \--- Plenoxels~\cite{plenoxel}, 
which boasts great 
visual effects, 
for representing the input scene.
While its simplicity and regular structure benefit patch-based algorithms,
important designs must be adopted.
Specifically,
we construct the exemplar pyramid via coarse-to-fine training Plenoxels on images of the input scene,
instead of trivially downsampling a pretrained high-resolution one.
Furthermore,
we transform the high-dimensional, unbounded, and noisy features of the Plenoxels-based exemplar at each scale into more well-defined and compact geometric and appearance features,
improving the robustness and efficiency in the subsequent patch matching.

On the other end,
we employ heterogeneous representations for the synthesis inside the generative nearest neighbor module.
Specifically,
the patch matching and blending operate in tandem at each scale to gradually synthesize an intermediate \emph{value}-based scene,
which will be eventually converted to a \emph{coordinate}-based counterpart at the end.
The benefits are several-fold:
a) the transition between consecutive generation scales, where the value range of exemplar features may fluctuate, is more stable;
b) the transformed features in the synthesized output is inverted to the original "renderable" Plenoxels features;
so that c) the visual realism in the Plenoxels-based exemplar is preserved intactly.
Last, 
working on voxels with patch-based algorithms necessarily leads to high computation issues.
So we use an \emph{exact-to-approximate} patch nearest-neighbor module in the pyramid,
which keeps the search space under a manageable range while introducing negligible compromise on the visual summary optimality.
These designs,
on a collective level,
essentially lay a solid foundation for an effective and efficient 3D generative model.

To our knowledge, 
our method is the \emph{first} 3D generative model that can generate 3D general natural scenes from a \emph{single} example,
with \emph{both} realistic geometry and visual appearance,
in large quantities and varieties.
We validate the efficacy of our method on random scene generation with an array of exemplars featuring a variety of general natural scenes,
and show the superiority by comparing to baseline methods.
The importance of each design choice is also validated.
Extensive experiments also demonstrates
the versatility of our method in several 3D modeling applications.

%% file: 2-related.tex
\section{Related Work}
\paragraph{3D Generative Models.}
The goal of 3D generative models is to synthesize 3D contents with realistic geometric structures and visual appearances.
While procedural models are capable of mass-producing particular 3D models,
they take expertise and time to obtain rules and elementary assets.
Hence,
automating this process has been an active area of research,
resulting in a vast body of work.

A prevalent route is the learning-based one, assuming having access to sufficient homogeneous samples for training.
Some learn to generate realistic 3D geometric structures, 
such as
indoor scenes~\cite{makeithome, wang2018deep, li2019grains, zhang2020deepindoor, wu2019datainterior, fu2017adaptiveindoor, ma2018language}, 
terrains~\cite{terrainreview, terraincnn},
urban scenes~\cite{feng2016crowd, nishida2016interactiveurban},
etc.
Others focus on the visual appearance, 
attempting to automatically texturize or assign materials for geometric scaffolds~\cite{kapp2020data, zhu2017indoorcolorization, jain2012materialmemex, kelly2018frankengan, hao2021gancraft}.
Another line has been directed at generating single objects with realistic structures or/and textures~\cite{scores, gao2021tm, chaudhuri2011probabilistic, li2017grass, li2021sp, ngp, park2018photoshape},
showing the potential in enriching the elementary asset library.
A dominant trend in recent has also emerged\cite{graf, giraffe, pi-gan, unconstrained, triplane, stylenerf}, 
where deep generative models are trained on large volumes of images collected from scenes of a specific category, 
to allow joint synthesis of realistic 3D structure and appearance with neural radiance fields.
Nevertheless, 
all these learning setups require large volumes of training data, and are limited in their ability to generalize, especially in terms of varied scene types.

A more relevant direction is the exemplar-based one, where one or a few exemplars featuring the scene of interest are provided.
However,
existing methods with algorithm designs tailored for certain scene types again draw clear boundaries of scene characteristics they can handle.
~\cite{zhou2007terrain} extract height field patches from exemplars to synthesize terrains, but the synthesis is guided with particular emphasis on dominant visual elements in terrains.
~\cite{merrell2008continuous, merrell2007example, merrell2010model} use structured units specified in the input exemplar to facilitate architecture model synthesis.
Extending texture image synthesis, 
~\cite{lee2011adaptive} synthesizes signed distance fields from an input geometry, 
the method can not generalize to complex general natural scenes, and the result is inadequate for displaying due to the lack of appearance properties.

In this paper, 
we aim for 3D general natural scenes, 
with an emphasis on generating both realistic geometry and appearance.
Lacking the necessary volume of data characterizing the target scene,
along with the difficulties of having ad hoc designs in presence of varying scene characteristics,
we advocate for synthesizing novel 3D scenes at the patch level, given a single exemplar scene.

\vspace{5px}
\noindent{\textbf{Generative Image Models.}}
Generative image models have made great strides in the past years.
State-of-the-art methods can now learn generative models from large volumes of homogeneous image samples,
achieving unprecedented success in producing realistic images~\cite{stylegan, stylegan2, vahdat2021score, dhariwal2021diffusion, kingma2018glow, vahdat2020NVAE}.
On the other end,
there has also been a surge of developments to learn a generative model from a single training image~\cite{singan, csg_singan, ingan, hinz2021improved}.
But, these learning-based single image models typically require a long training time.
Differing from these learning-based paradigms,
a classical patch-based approach, that dates back long before the deep learning era, is revived in~\cite{drop, vgpnn, elnekave2022generating},
showing amazing performance.
The core of these models is to maximize the bidirectional patch similarity between the input and synthesized output in a coarse-to-fine manner,
and have demonstrated their capability to generate diverse outputs from a single exemplar image, with orders of magnitude faster than learning-based ones.
Our work is particularly inspired by this line of work but must address challenges arising from lifting the multi-scale generative patch-based framework to \emph{effective} and \emph{efficient} 3D scene generation.

\vspace{5px}
\noindent{\textbf{3D Scene Representations.}}
While it is common to represent an image as a distributed amplitude of colors over a 2D grid, 
more often than not, the 3D representation varies.
Polygon meshes and points offer a compact representation, with precedents in patch-based synthesis~\cite{sharf2004context, harary2014context}, 
but the irregularity makes them intractable for high-quality 3D generation.
The same holds for point clouds.
Recently, 
the community has indeed witnessed a revolution started by an emerging representation, 
i.e.\ neural radiance field~\cite{nerf}, 
which approximates the 5D plenoptic function~\cite{plenoptic} of the underlying scene with neural networks and shows unprecedentedly photo-realistic visual results.
An explosion of techniques occurred since then that improves the representation in various aspects~\cite{svs, fvs, idr, yariv2021volume, plenoctrees, hedman2021snerg, instantngp, sun2022direct}. 
We refer readers to~\cite{tewari2020state, xie2021neural} for more in-depth summaries.
Among these variants, 
we opt for a simple yet expressive voxel-based representation \--- Plenoxels~\cite{plenoxel},
which has shown great competence on novel view synthesis.
Its simplicity and regular structure benefit patch-based algorithms,
however,
important designs must be taken to fit it into our framework for high-quality generation of general natural scenes.

\vspace{5px}
\noindent{\textbf{Concurrent Work.}}
Concurrent works~\cite{singrav, tinGAN} propose to learn a 3D generative model from images of an input scene, producing variations that can be rendered with realistic imagery.
~\cite{wu2022learning} focus on generating diverse geometric structures from an input shape.
Their core idea is to extend 2D SinGAN~\cite{singan} for learning the internal distribution in the 3D exemplar, 
differing significantly from our technical route.
While these methods require a long training time (typically days),
our method can generate high-quality samples in minutes, without offline training.
Last, ~\cite{poole2022dreamfusion} can generate arbitrary 3D models represented by NeRF, with pretrained powerful image diffusion models as priors.

%% file: 3-method.tex
\section{Method}
\input{math_utils}

\begin{figure}[t!]
    \centering
    \includegraphics[width=1.\linewidth]{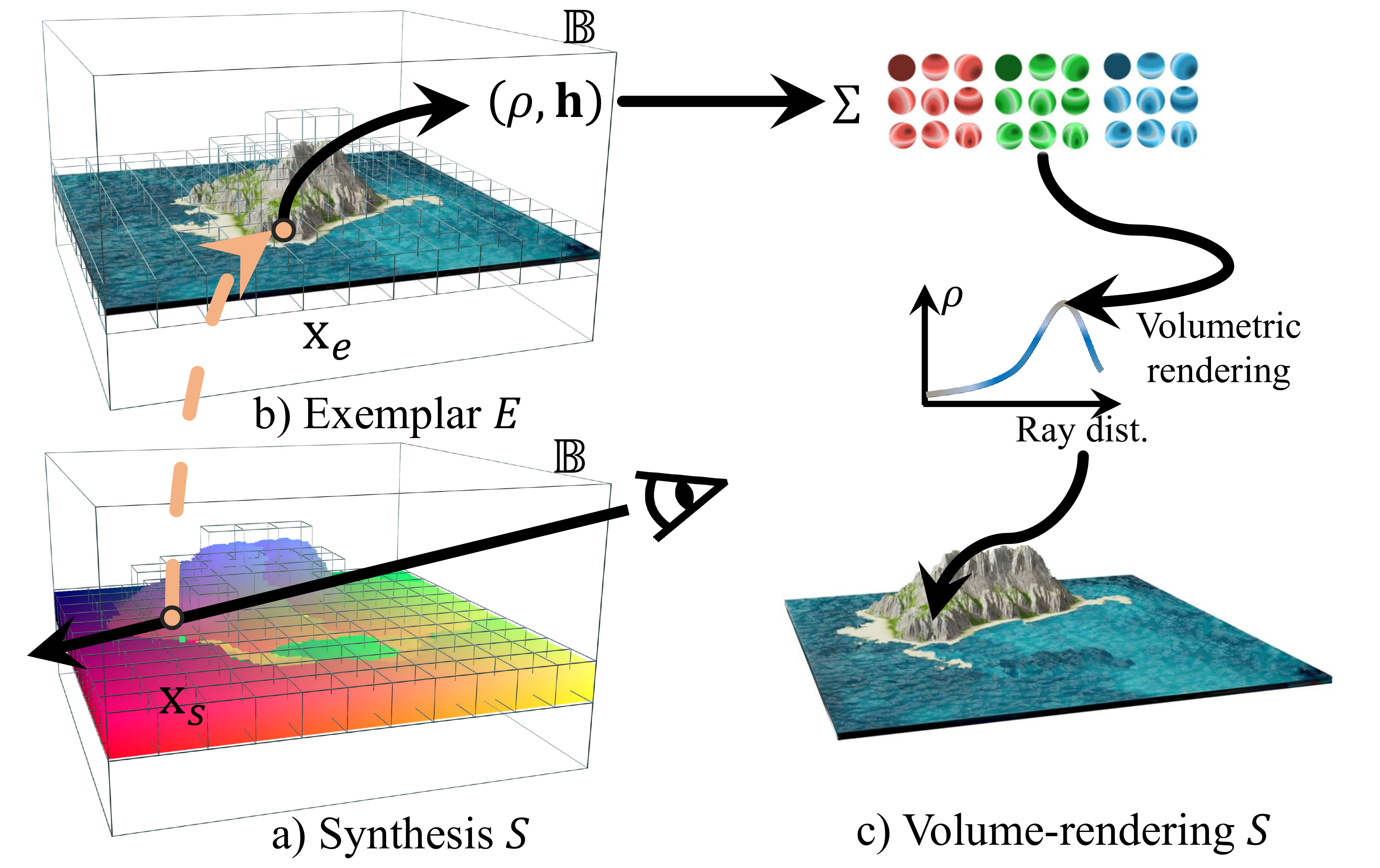}
    \caption{
    a) The synthesized scene $\synthesis$ is represented as a field mapping a coordinate in $\synthesis$ to one in $\exemplar$.
    b) The Plenoxels-based exemplar $\exemplar$ uses a sparse grid, where each occupied voxel stores a scalar opacity $\density$ and spherical harmonic coefficients $\appearance$.
    c) Appealing imagery of $\synthesis$ can be produced via the volumetric rendering function.
    Empty voxels are omitted for simplicity. 
    }
    \label{fig:representation}
\end{figure}

The input 3D scene to our method can be a real-world or digital scene,
as we first train Plenoxels on the images of the input scene to obtain a Plenoxels-parameterized exemplar.
Then,
our method synthesize novel variations at the patch level, with a multi-scale generative patch-based framework.
In the following,
we describe important designs w.r.t. the scene representation (Section~\ref{sec:representation} \&~\ref{sec:framework}) and the generative patch nearest-neighbor field module (Section~\ref{sec:nnf}), that, integrated into the multi-scale patch-based framework (Section~\ref{sec:framework}), contribute collectively to our success. 

\subsection{Scene Representations}
\label{sec:representation}
\paragraph{Exemplar Scene Representation.}
We assume the exemplar scene $\exemplar$ lies within an axis-aligned box $\bound$ centered at the origin, around which we can distribute cameras to capture images for training Plenoxels.
As per Plenoxels, $\exemplar$ is represented by a sparse voxel grid, 
where each occupied voxel center stores features including a scalar opacity $\density$ and a vector of spherical harmonic (SH) coefficients $\appearance$ for each color channel:
 $  \exemplar: \coord \to (\density, \appearance),   $
where $\coord$ indicates a voxel center within $\bound$.
These features can be further trilinearly interpolated to model the full plenoptic function continuously in space. 
Notably,
the appearance feature uses 2-degree harmonics, 
which requires 9 coefficients per color channel for a total of 27 harmonic coefficients per voxel.

\vspace{5px}
\noindent{\textbf{Exemplar Transformation.}}
While Plenoxels features can be used to render pleasing imagery,
naively using them for the patch distance is unsuitable.
Density values are not well-bounded, contain outliers, and can not accurately describe the geometric structure within a patch.
On the other hand,
high-dimensional SH coefficients are excessively consumptive for patch-based frameworks.
Hence, we transform the exemplar features for the input to the generative patch nearest neighbor module.
First,
the density field is converted to a signed distance field (SDF).
Specifically,
the signed distance at each voxel is computed against the surface mesh extracted from the density field by Marching Cubes~\cite{marchingcubes}. 
Note that Plenoxels prunes unnecessary voxels during training, 
which creates holes and irregular structures in invisible regions.
So we flood-fill these regions with high-density values, prior to the mesh extraction.
Last,
we rescale and truncate the signed distance to ignore distance values far away from the surface.
Formally, the geometry transformation is as follow:
$\tsdf(\coord) = \max \Bigl( {-1}, \min \bigl( 1, \sdf(\coord) / t \bigr) \Bigr)$,
where the truncated scale $t$ is set to 3 times of the voxel size at each generation scale.
Moreover, we normalize SH coefficient vectors and use the principal component analysis (PCA) to reduce the dimensionality (from 27 to 3 by default), 
significantly reducing the computation overhead.
Finally,
the transformed exemplar $\texemplar$ is now given as:
\begin{equation}
\begin{gathered}
   \texemplar: \coord \to \big( \tsdf(\coord), \pca(\appearance) \big), 
\end{gathered} 
\end{equation}
where $\tsdf(\cdot)$ denotes transforming of the geometric feature, and $\pca(\cdot)$ transforming the appearance feature.

\vspace{5px}
\noindent{\textbf{Synthesized Scene Representation.}}
In the multi-scale generation, 
the output scene $\synthesis$ at each scale is represented by a coordinate-based mapping field, 
instead of a value-based one that stores features. 
Specifically,
$\synthesis$ is represented as a field that maps a 3D voxel center in the synthesis grid to one in the exemplar $E$, 
 $ \synthesis: \coord_s \to \coord_e $,
with which the original Plenoxels features $\exemplar\big(\synthesis(\coord_s)\big)$ can be queried for $\synthesis$.

Note, in addition to discrete grid samplings, 
dense samplings $\coord_s$ in $\synthesis$ can also be mapped to the continuous exemplar space, 
by simply considering the local offset $\delta$ to the nearest voxel center,
i.e., 
$ \synthesis(\coord) = \synthesis ( \mathbf{N}(\coord) ) + \delta  $,
where $\mathbf{N}(\cdot)$ returns the nearest voxel center of $\coord$.
This is particularly useful, 
as it enables upscaling $\synthesis$ to finer grids in the multi-scale framework, and sufficient sampling for rendering the final generation result with high-quality imagery.

\vspace{5px}
\noindent{\textbf{Viewing Synthesized Results.}}
The synthesized scene can be projected onto 2D through the volume rendering equation as in NeRF~\cite{nerf}, yielding highly photo-realistic imagery under varying views.
We refer readers to~\cite{plenoxel} for more details.
Figure~\ref{fig:representation} illustrates how a synthesized result, paired with the exemplar, can display appealing imagery.

\subsection{Multi-scale Generation}
\label{sec:framework}
\begin{figure}[t!]
    \centering
    \includegraphics[width=1.\linewidth]{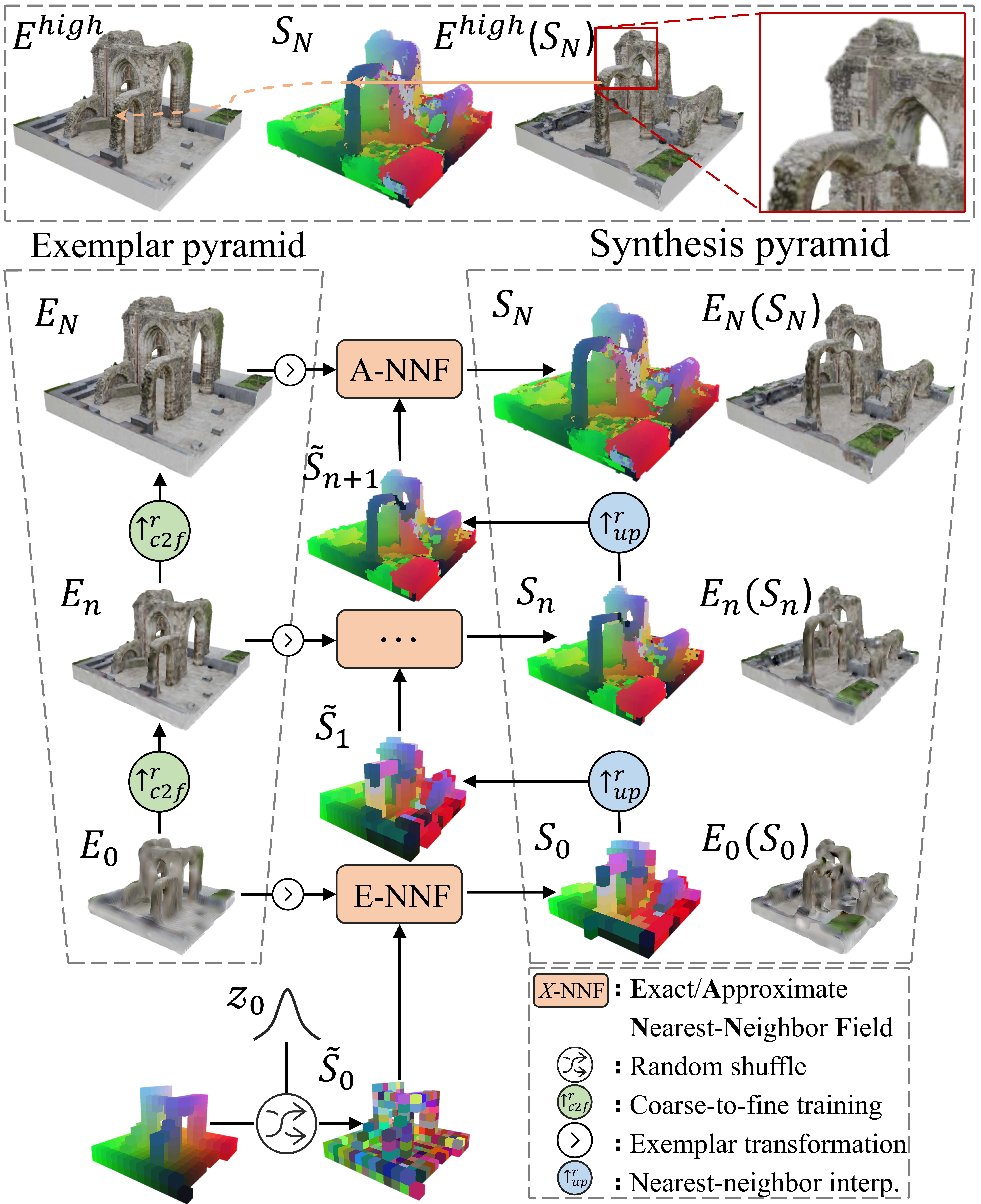}
    \caption{
    Multi-scale generation.
    At each scale $n$, the NNF module updates the generation based on the transformed exemplar $\texemplar_n$ and an initial guess $\synthesisup_n$ upscaled from the previous. 
    The coarsest scale takes a shuffled identity mapping field as input.
    Note that our coordinate-based representation $\synthesis_N$ can map to patches in a higher-resolution exemplar for higher quality (top row).
    }
    \label{fig:multi-scale}
\end{figure}
We use the same multi-scale framework as in previous works~\cite{efros2001image, drop, singan}, which generally employs a coarse-to-fine process,
so we have the opportunity to synthesize a more detailed scene based on an initial guess upscaled from the previous scale.
In this pyramidal pipeline, 
different information is captured and reproduced at varying scales, spanning from global layouts at coarser scales to fine geometric and appearance details at finer scales (See Figure~\ref{fig:multi-scale}).

\begin{figure}[t!]
    \centering
    \includegraphics[width=1.\linewidth]{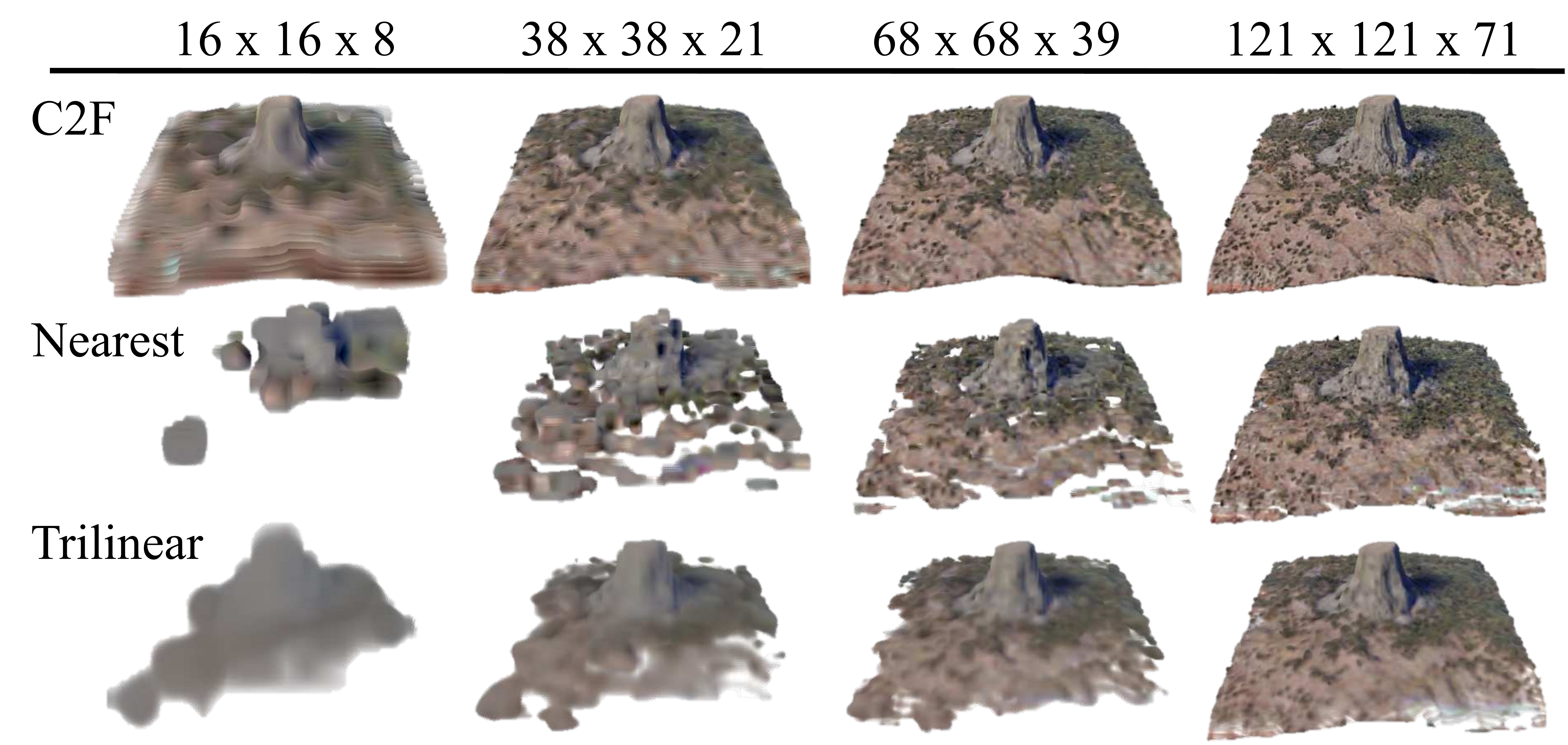}
    \caption{
    Exemplar pyramid.
    Coarse-to-fine training (top) shows more consistency between consecutive exemplars,
    whereas common downsampling algorithms (mid and bottom) result in missing geometry (e.g., the ground) and blurry appearance.
    }
    \label{fig:pyramid_construction}
\end{figure}

\paragraph{Exemplar Pyramid Construction.}
Given the input scene,
we build a pyramid $(\exemplar_0, ... , \exemplar_N)$, 
where $\exemplar_{n-1}$ is a downscaled version of $\exemplar_n$ by a factor $r^{-1}$ ($r = 4/3$).
By default, we use $N=7$ (8 scales in total) for balancing quality and efficiency. 
Specific resolutions in the pyramid are listed in the supplementary.
When working with an exemplar pyramid obtained by recursively downsampling a pretrained high-resolution exemplar,
we observed lots of artifacts due to missing features at coarser exemplars,
and
severe feature inconsistency between exemplars at consecutive scales.
Hence, we build the exemplar pyramid by coarse-to-fine training Plenoxels,
at increasing resolutions synchronized with the multi-scale framework.
Such exemplar pyramid prevents losing thin structures at coarser scales, and offers rather smoother transition and consistent features between consecutive exemplars, 
leading to stable transition in the multi-scale generation (See Figure~\ref{fig:pyramid_construction}).

\paragraph{Coarse-to-fine Generation.}
At each scale $n$, 
an initial guess $\synthesisup_{n}$ is produced by upsampling the output in the previous scale: $\synthesisup_{n} = \synthesis_{n-1}\uparrow^{r}$, with the same factor $r$ to match with the exemplar.
Then, the mapping field in $\synthesisup_{n}$ is updated by the generative nearest neighbor module with matched coordinates in the exemplar.
The patch size at all scales is $p^3$ ($p=5$ by default), which captures around 1/3 of the content in the coarsest exemplar.
Unlike adding noise to raw exemplar values as in the image synthesis, 
our initial guess $\synthesisup_0$ at the coarsest scale is an identity mapping field shuffled with Gaussian noise $z_0 = \mathcal{N}(0, \sigma^2)$ , $\sigma=0.5$ by default, scaled by the extents of the bounding box $\bound$, which is natural for our coordinate-based synthesis.

\subsection{Generative Patch Nearest-Neighbor Field}
\label{sec:nnf}

Usually, two stages, namely the patch \emph{matching} and \emph{blending}, 
operate in tandem in the nearest neighbor field (NNF) search of patch-based algorithms.
Specifically,
the matching finds the most suitable patch from the exemplar for each in $\synthesis$, 
and then the latter blends of the \emph{values} of overlapping patches together.
This is vital to a robust EM-like optimization in patch-based image synthesis~\cite{patchmatch, barnes2017survey},
leading to converging synthesis results in several iterations.

\begin{figure*}[th!]
  \centering
  \includegraphics[width=\linewidth]{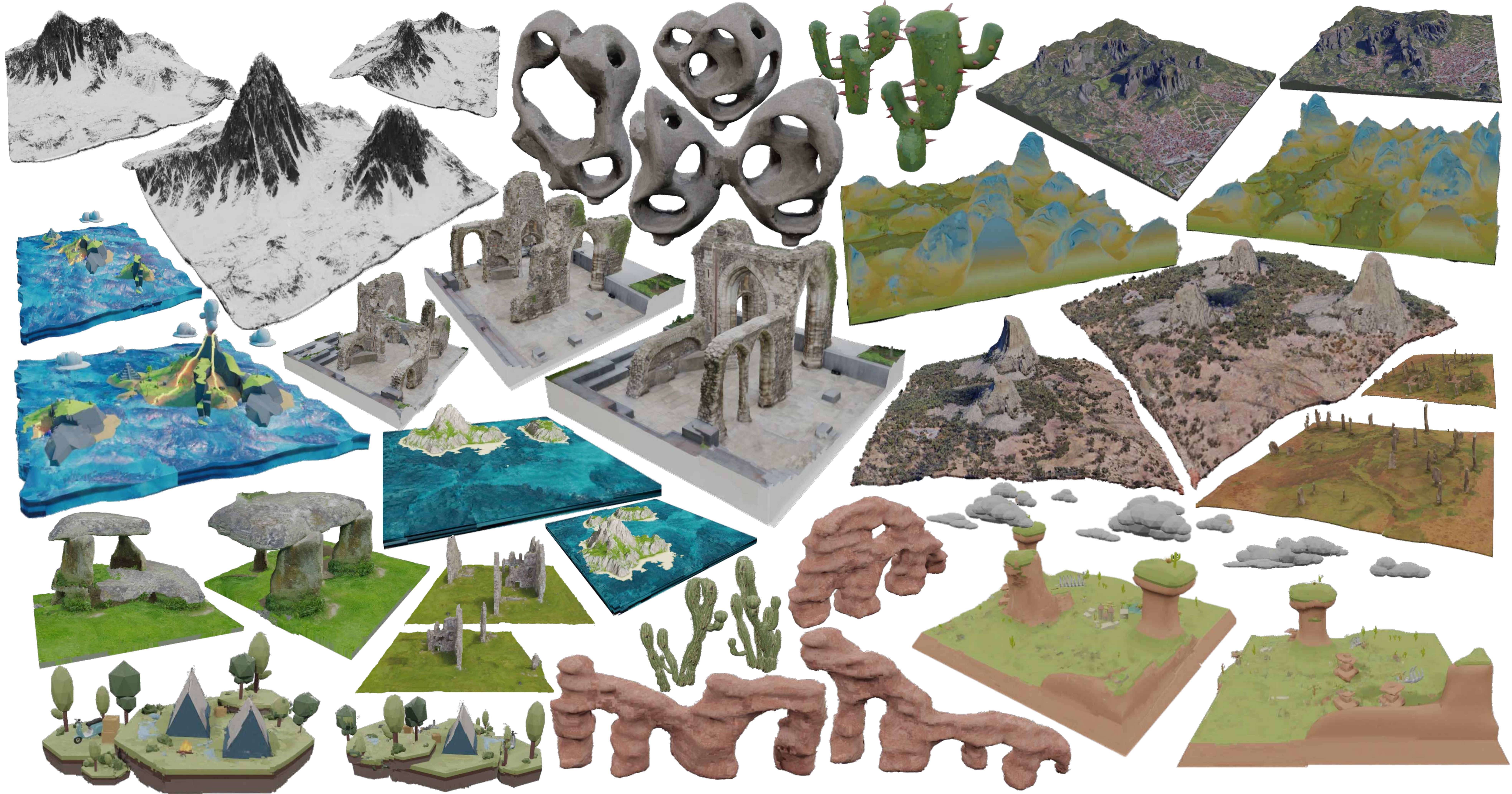}
  \caption{
  Random generation.
  Our method generalizes to all these scenes with highly varied structures and appearances,
  producing highly diverse and realistic scenes.
  The supplementary presents exemplars and more artistic imagery rendered with these 3D scenes.
  }
  \label{fig:result_gallery}
\end{figure*}

\paragraph{Value-/Coordinate-based Synthesis.}
In this work,
we use heterogeneous representations for the synthesis in NNF.
Specifically,
at each scale $n$,
the patch matching and blending first operate in tandem for $T-1$ iterations, to gradually synthesize an intermediate \emph{value}-based scene with averaged values over overlapping patches.
Then, when the synthesis is stable at the last iteration, the final output of NNF uses coordinate-based representation, which stores only the center location of the nearest patch in $\texemplar_n$.
As aforementioned,
this design offers stable transition between consecutive generation scales, where the value range of exemplar features may fluctuate,
and, importantly,
helps us trace back to the original Plenoxels features that can be rendered into photo-realistic imagery, via simply mapping to the original exemplar, even to a higher-resolution version for the final generated scene (See top of Figure~\ref{fig:multi-scale}).
Specifically, 
each iteration in NNF at each scale proceeds as follows:

(1) \textit{Extract Patches}: 
  Patches in $\texemplar_n(\synthesisup_n)$ are extracted to form a query patch set $\qset$, and ones in $\texemplar_n$ form a key set $\kset$.
  
(2) \textit{Match Nearest Neighbors}: 
  We first compute distance between each query patch $\qset_i$ and each key patch $\kset_j$ as the weighted sum of the appearance and geometric features using $L2$ distance: 
   \begin{equation}
   \begin{gathered}
       \dist_{i,j} = {w_a} || \qset_{i,j}^{a} - \kset_{i,j}^{a} ||^2 + (1-{w_a}) || \qset_{i,j}^{g} - \kset_{i,j}^{g} ||^2 ,
   \end{gathered} 
   \end{equation}
  where $w_a$ (0.5 by default) is the trade-off parameter.
  To control the visual completeness in the synthesis by the bidirectional similarity \cite{simakov2008summarizing}, the final patch similarity scores normalize the distance with a per-key factor: 
   \begin{equation}
   \begin{gathered}
       \simi_{i,j} =  \frac{\dist_{i,j}}{(\alpha + \min_{l}(\dist_{l,j}))} ,
   \end{gathered} 
   \end{equation}
  where $\alpha$ (0.01 by default) controls the degree of completeness, and smaller $\alpha$ encourages completeness.
  
(3) \textit{Update $\synthesisup_n$}: 
For each query patch $\qset_i$ in $\texemplar_n(\synthesisup_n)$, we find its nearest patch in $\kset_l$, then update $\synthesisup_n$ with averaged values over overlapping patches for the first $T-1$ iterations,
and with the nearest patch center for the last iteration.

\paragraph{Exact-to-Approximate NNF.}
Although the computation above can be in parallel performed on GPUs, 
brutally enumerating all pairs of patches would apparently lead to surprisingly huge distance matrices as the resolution increases, 
preventing us from obtaining high-resolution synthesis even with modern powerful GPUs. 
Hence, to avoid searching in tremendous space, 
we propose to perform the NNF in an \emph{exact-to-approximate} manner.
Specifically,
at first 5 coarser scales, \emph{exact nearest-neighbor field} (E-NNF) search is performed with $T_e = 10$ times to stabilize global layout synthesis when the memory consumption is low.
At rest 3 finer scales, an \emph{approximate nearest-neighbor field} (A-NNF) search \--- PatchMatch\cite{patchmatch} \--- with jump flood ~\cite{jumpflooding} is used for $T_a = 2$ times to reduce memory footprint from $O(M^2)$ to $O(M)$ ($M$ is the number patches), 
which is equivalent to only considering visual coherence.

%% file: math_utils.tex
\newcommand{\exemplar}{ E }
\newcommand{\texemplar}{ \hat{E} }
\newcommand{\synthesis}{ S }
\newcommand{\synthesisup}{ \tilde{S} }
\newcommand{\bound}{ \mathbb{B} }

\newcommand{\coord}{ \textbf{x} }
\newcommand{\density}{ \rho }
\newcommand{\appearance}{ \textbf{h} }
\newcommand{\rgb}{ \textbf{c} }
\newcommand{\ray}{ \textbf{r} }
\newcommand{\view}{ \textbf{d} }
\newcommand{\expo}{ \text{exp} }

\newcommand{\sdf}{ SDF }
\newcommand{\tsdf}{ G }
\newcommand{\pca}{ P }

\newcommand{\qset}{ Q }
\newcommand{\kset}{ K }

\newcommand{\dist}{ D }
\newcommand{\simi}{ C }

\newcommand{\nbgen}{ M }

%% file: 4-exp.tex
\section{Experiments}
\label{sec:exp}
We collected a rich variety of 3D scene models to examine the performance of our method on random scene generation,
ranging from 
rocks to plants, sculptures, landscapes, terrains, artistic scenes, etc.
Some are digitalized \emph{real-world} scenes, e.g., the \emph{Devil's Tower}.
These scenes possess varying degrees of complexity in terms of geometry and appearance.
In the following, we present experiments conducted to evaluate various aspects of the proposed solution.
Unless specified,
we use the default parameters described above,
512 for the resolution along the max dimension of the $\exemplar^{high}$, 
and $512 \times 512$ image resolution for rendering.
Full visualization of all exemplars,
more technical details and experimental results can be found in the supplementary.

\begin{figure*}[t!]
  \centering
  \includegraphics[width=\linewidth]{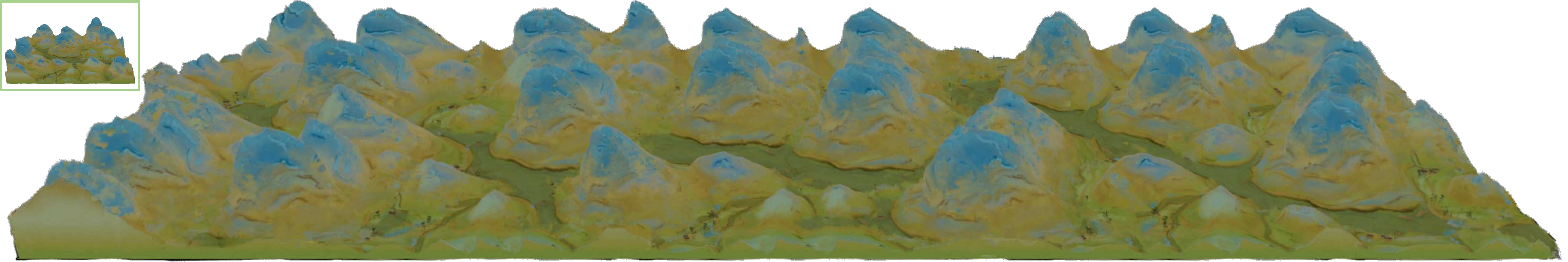}
  \caption{
  A novel \emph{"A Thousand Li of Rivers and Mountains"}~\cite{wangximeng} is rendered from a generated 3D sample, that is of a different size, resolution and aspect ratio to the \emph{Vast Land} exemplar (inset).
  Specification:
  $\exemplar_N$ - $288\times288\times112$,
  $\exemplar^{high}$ - $512\times512\times200$,
  $\synthesis_N$ - $747\times288\times112$,
  $\exemplar^{high}(\synthesis_N)$ - $1328\times512\times200$,
  final rendering resolution - $4096\times1024$.
  }
  \label{fig:High-reso-tiling}
\end{figure*}

\begin{figure}[t]
    \centering
    \includegraphics[width=\linewidth]{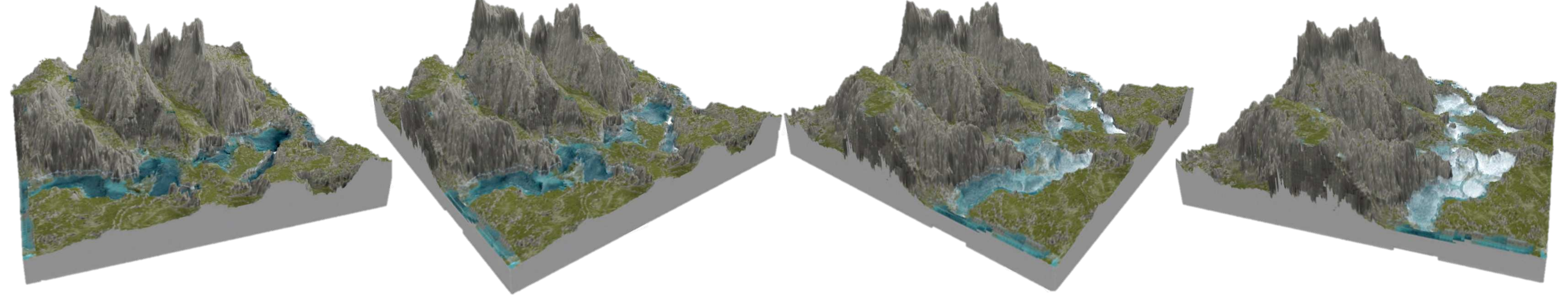}
    \caption{
    View-dependent effects in our synthesized results.
    See the reflection on the river changing under spinning cameras. 
    }
    \label{fig:view_effects}
\end{figure}

\vspace{5px}
\noindent{\textbf{Random Generation.}}
Figure~\ref{fig:result_gallery} presents results obtained by our method on exemplar-based random scene generation.
These results show our method can generalize to scenes of highly varied features, 
yielding high-quality and diverse scenes similar to the exemplar.
A particular feature of our method is the photo-realism and view-dependent effects of the exemplar are inherited in the results,
as evidenced by Figure~\ref{fig:result_gallery} and~\ref{fig:view_effects}.
Each sample is generated in 1$\sim$3 minutes on a V100 GPU depending on the scenes, and viewing the results can be executed at an interactive rate (15 fps).

\begin{figure}[b!]
    \centering
    \includegraphics[width=\linewidth]{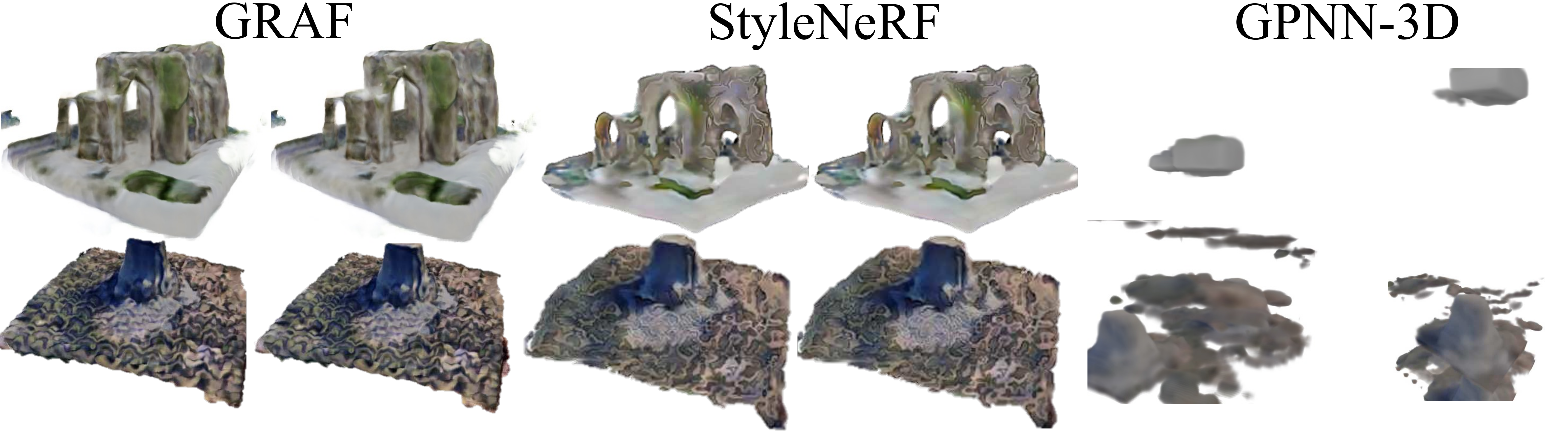}
    \caption{
    Visual comparisons.
    GAN-based baselines 
    suffer from severe mode collapse, 
    producing samples (two shown) almost identical to the input.
    GPNN-3D fails on the task.
    }
    \label{fig:baseline_visuals}
\end{figure}

\vspace{5px}
\noindent{\textbf{Comparisons.}}
We particularly compare to GRAF and StyleNeRF, which are 
representative GAN-based 
3D generative models.
We cast them into exemplar-based models
via training separately on images of each exemplar.
In addition,
we also compare to GPNN-3D, which trivially extends~\cite{drop} for our task.
We investigate the advantages of exemplar-based scene generation using our method against these alternatives,
on various exemplars listed in Table~\ref{tab:comparison_table}.
Figure~\ref{fig:baseline_visuals} presents part of their visual results.
Generally, 
GAN-base baselines suffer from notorious mode collapse, 
producing almost identical results due to lacking diverse training scenes.
The visuals also tend to be more blurry and noisy, compared to our sharp imagery.
GPNN-3D can not synthesize high-resolution results due to computational efficiency issues, 
and quickly fails at coarse scales, producing meaningless content.
For quantitative comparisons,
we produce 
50
generated scenes from each exemplar with each method,
render multi-view images and extract 3D surface points of the exemplar and of each generated scene,
and then rate the \emph{Visual Quality} (V-Qua.) 
\emph{Visual Diversity} (V-Div.), \emph{Geometry Quality} (G-Qua.), and \emph{Geometry Diversity} (G-Div.) using common metrics employed in both 2D~\cite{singan} and 3D~\cite{mmsc} generation.
The supplementary contains more details.
Table~\ref{tab:comparison_table} presents quantitative results,
where, by rating with the combination of these established metrics, 
ours outperforms baselines by large margins, suggesting high quality and diversity from both 2D and 3D perspective.
\input{table/4-exp-comparison_table.tex}

\begin{figure}[t!]
    \centering
    \includegraphics[width=\linewidth]{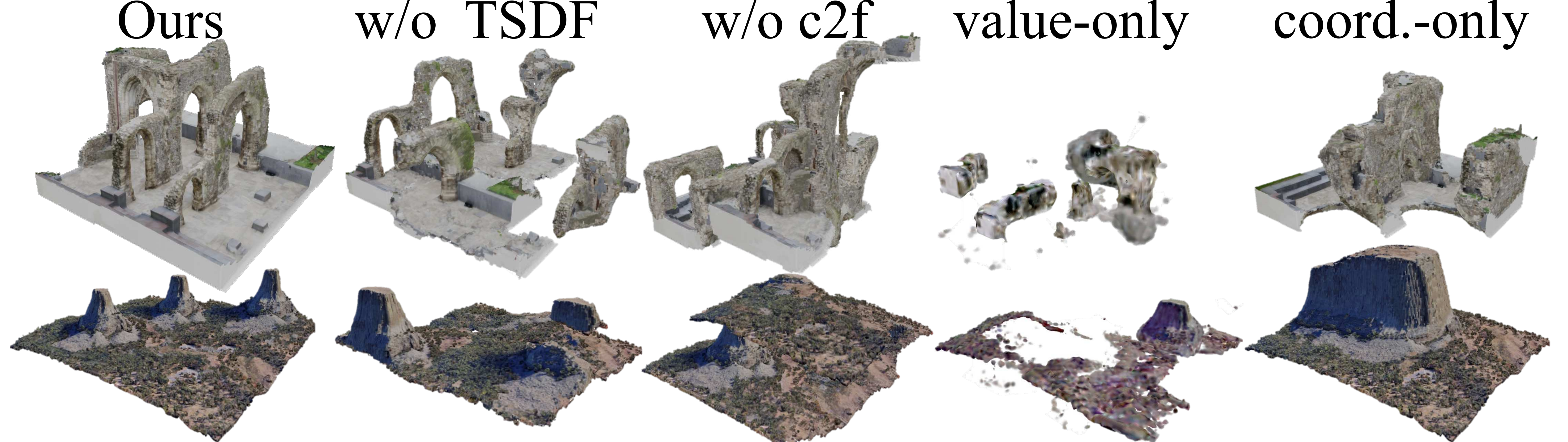}
    \caption{
    Ablation study. 
    Ours (w/o TSDF) and (w/o c2f) can not well preserve the geometric structures.
        Ours (value-only) fails and produces with noisy content,
    while Ours (coord.-only) is unstable, easily leading to bulky structures or holes.
    }
    \label{fig:ablation_visuals}
\end{figure}

\vspace{5px}
\noindent{\textbf{Ablation.}}
We compare to several variants derived from our full method:
1) \emph{Ours (w/o TSDF)} uses an occupancy field, instead of TSDF, converted from the exemplar density field for geometric features; %
2) \emph{Ours (w/o c2f)} drops the deep coarse-to-fine exemplar training, and instead recursively trilinearly interpolates a high-resolution exemplar;
3) \emph{Ours (value-only)} uses only value-based synthesis in NNF, and does not use TSDF and PCA as we can not trace back to original Plenxels features, and the maximum resolution is limited to 68;
4) \emph{Ours (coord.-only)} uses only coordinate-based synthesis in NNF.
Figure~\ref{fig:ablation_visuals} and Table~\ref{tab:ablation_table} present the qualitative and quantitative comparison results, respectively,
showing the importance of each algorithmic design.

\input{table/4-exp-ablation_table.tex}

\vspace{5px}
\noindent{\textbf{Higher-resolution Generation.}}
1) 
In Figure~\ref{fig:High-reso-tiling},
we show that our method supports generating a result scene of different size to the exemplar, and particularly of a much higher resolution and different aspect ratio.
See specifications in the caption.
2)
In addition,
we also stress test with a very high-resolution setting,
where $\exemplar_N$ has 288 voxels along the max dimension, and our method can still synthesize a highly plausible sample in $\sim$10 minutes.
We observed slightly improved visual quality over the default setting, 
as the default is sufficient for most complicated scenes.
Results and details can be found in the supplementary.

%% file: table/4-exp-comparison_table.tex
\begin{table*}[t!]
\caption{
Quantitative comparisons.
Ours outperforms baselines by large margins, with high quality and diversity scores in terms of both visual and geometric content. 
We highlight top two in bold and underline the top one.
Note GPNN-3D's high diversity scores can be explained by noisy contents shown in the visual results.
}
\scriptsize
\centering
\setlength{\tabcolsep}{2.pt}

\begin{tabular}{c|cccc|cccc|cccc|cccc} 
\toprule
 & \multicolumn{4}{c|}{GRAF} & \multicolumn{4}{c|}{StyleNeRF} & \multicolumn{4}{c|}{GPNN-3D} & \multicolumn{4}{c}{Ours} \\ 
\midrule
 & V-Qua.$\downarrow$ & V-Div.$\uparrow$ & G-Qua.$\downarrow$ & G-Div.$\uparrow$ & V-Qua.$\downarrow$ & V-Div.$\uparrow$ & G-Qua.$\downarrow$ & G-Div.$\uparrow$ & V-Qua.$\downarrow$ & V-Div.$\uparrow$ & G-Qua.$\downarrow$ & G-Div.$\uparrow$ & V-Qua.$\downarrow$ & V-Div.$\uparrow$ & G-Qua.$\downarrow$ & G-Div.$\uparrow$ \\
St Alphage & \textbf{0.078} & \textbf{0.046} & \textbf{\uline{0.473}} & 0.040 & 0.206 & 0.032 & 0.769 & 0.012 & 3.929 & 0.041 & 134.997 & 0.059 & \textbf{\uline{0.022}} & \textbf{\uline{0.312}} & \textbf{0.612} & \uline{\textbf{0.473}} \\
Devil's Tower & \textbf{0.233} & 0.084 & \textbf{0.480} & 0.021 & 0.470 & 0.021 & 1.000 & 0.011 & 2.495 & \textbf{\uline{0.694}} & 6.545 & \textbf{\uline{1.974}} & \textbf{\uline{0.032}} & \textbf{0.203} & \uline{\textbf{0.304}} & \textbf{0.207} \\
Desert Lowpoly & \textbf{0.057} & 0.048 & 0.721 & 0.043 & 0.255 & 0.026 & 0.813 & 0.018 & 1.312 & \textbf{\uline{0.405}} & \textbf{\uline{0.344}} & \textbf{\uline{1.048}} & \textbf{\uline{0.020}} & \textbf{0.312} & \textbf{0.568} & \textbf{0.454} \\
Green Island & \textbf{0.294} & 0.047 & \textbf{0.277} & 0.015 & 0.606 & 0.015 & 0.669 & 0.014 & 0.254 & \uline{\textbf{1.136}} & 18.228 & \uline{\textbf{17.673}} & \uline{\textbf{0.044}} & \textbf{0.172} & \uline{\textbf{0.097}} & \textbf{0.081} \\
Stone Arch & 0.101 & 0.055 & 0.603 & 0.029 & \textbf{0.060} & 0.011 & \textbf{0.339} & 0.005 & 1.608 & \textbf{\uline{0.504}} & 53.943 & \textbf{\uline{29.448}} & \textbf{\uline{0.003}} & \textbf{0.146} & \textbf{\uline{0.126}} & \textbf{0.100} \\
Mountain & 0.410 & 0.060 & \textbf{\uline{0.498}} & 0.022 & \textbf{0.222} & 0.037 & \textbf{0.757} & 0.010 & 2.602 & \textbf{\uline{0.787}} & 5.947 & \textbf{\uline{1.674}} & \textbf{\uline{0.105}} & \textbf{0.391} & 0.935 & \textbf{0.467} \\
Vast Land & \textbf{0.072} & 0.058 & 0.994 & 0.229 & 0.219 & 0.023 & 1.047 & 0.017 & 0.907 & \textbf{\uline{0.348}} & \textbf{0.690} & \textbf{\uline{1.840}} & \textbf{\uline{0.014}} & \textbf{0.124} & \textbf{\uline{0.177}} & \textbf{0.105} \\
\bottomrule
\end{tabular}

\label{tab:comparison_table}
\end{table*}%

%% file: table/4-exp-ablation_table.tex
\begin{table}[t!]
\caption{
Quantitative ablation results.
While some variants produce higher diversity scores with meaningless noisy contents,
ours consistently produce \emph{diverse} results with \emph{highest} quality scores. 
}
\small
\setlength{\tabcolsep}{1.5pt}
\centering

\begin{tabular}{c|c|cccc} 
\toprule
\multicolumn{1}{c}{} &  & V-Qua.$\downarrow$ & V-Div.$\uparrow$ & G-Qua.$\downarrow$ & G-Div.$\uparrow$ \\ 
\midrule
\multirow{5}{*}{St Alphage} & Ours & \textbf{\uline{0.022}} & 0.312 & \textbf{\uline{0.612}} & 0.473 \\
 & w/o TSDF & 0.114 & \textbf{\uline{0.568}} & 1.176 & \textbf{1.105} \\
 & w/o c2f & \textbf{0.024} & \textbf{0.353} & \textbf{0.847} & 0.639 \\
 & value-only & 3.779 & 0.054 & 56.304 & \textbf{\uline{3.823}} \\
 & coord.-only & 0.044 & 0.336 & 1.003 & 0.719 \\ 
\midrule
\multirow{5}{*}{Devil's Tower} & Ours & \textbf{\uline{0.032}} & 0.203 & \textbf{0.304} & 0.207 \\
 & w/o TSDF & 0.047 & 0.263 & 0.422 & 0.350 \\
 & w/o c2f & 0.082 & \textbf{\uline{0.547}} & 2.101 & \textbf{3.500} \\
 & value-only & 1.795 & \textbf{0.344} & 14.122 & \textbf{\uline{7.650}} \\
 & coord.-only & \textbf{0.041} & 0.201 & \textbf{\uline{0.256}} & 0.492 \\ 
\midrule
\multirow{5}{*}{Desert Lowpoly} & Ours & \textbf{\uline{0.020}} & 0.312 & \textbf{\uline{0.568}} & 0.454 \\
 & w/o TSDF & \textbf{0.041} & \textbf{0.462} & \textbf{1.347} & 1.007 \\
 & w/o c2f & 0.049 & 0.457 & 1.745 & 1.097 \\
 & value-only & 0.763 & 0.419 & 29.047 & \textbf{\uline{6.674}} \\
 & coord.-only & 0.100 & \textbf{\uline{0.487}} & 2.754 & \textbf{1.526} \\
\bottomrule
\end{tabular}

\label{tab:ablation_table}
\end{table}%

%% file: 5-application.tex
\vspace{5px}
\noindent{\textbf{Applications.}}
\begin{figure}[t!]
    \centering
    \includegraphics[width=\linewidth]{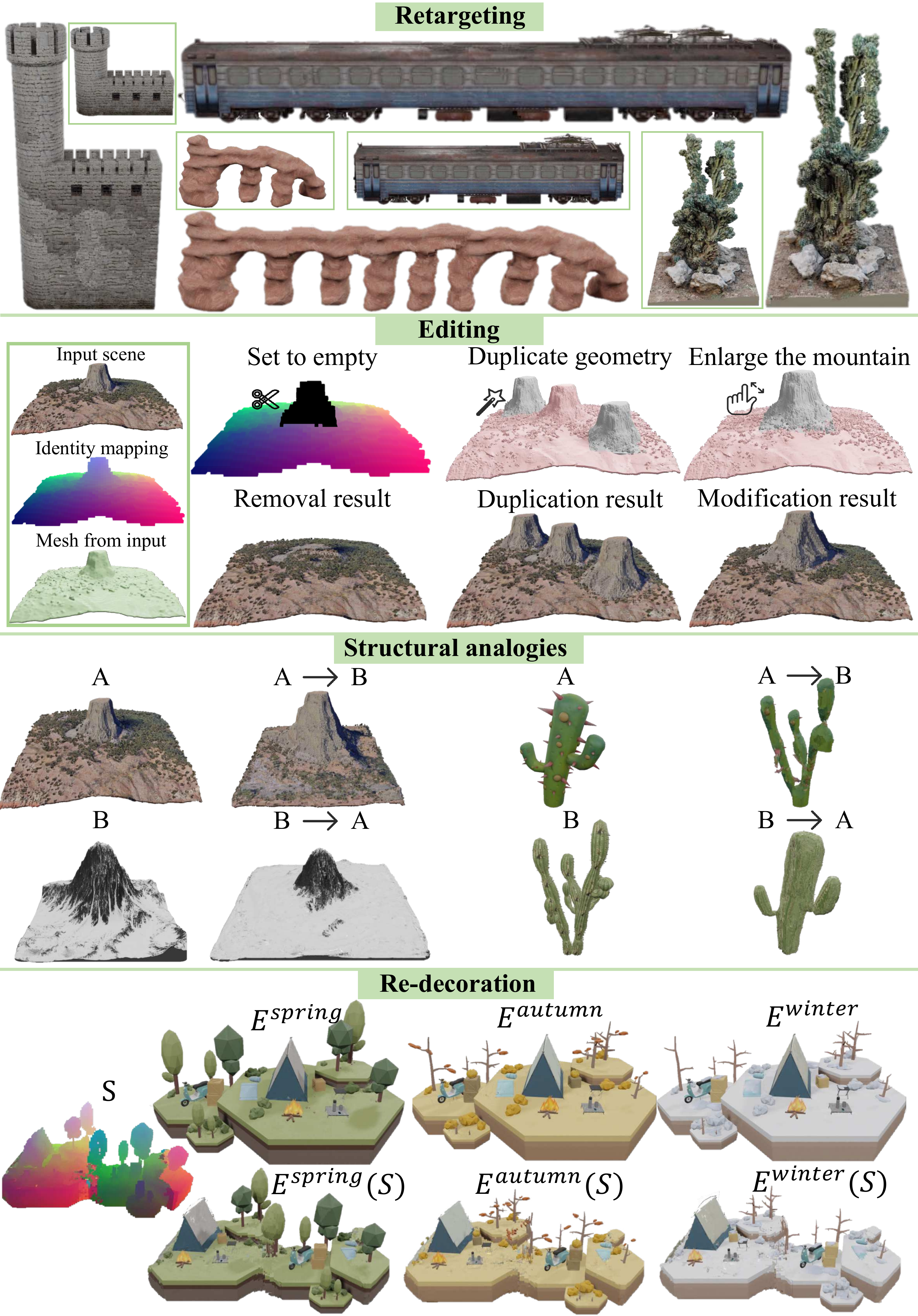}
    \caption{
    Applications.
    \nth{1}: Retargeting 3D scenes (marked in boxes).
    \nth{2}: Editing a 3D scene (removal, duplication and modification).
    \nth{3}: Structural analogies.
    A $\to$ B $=$ Visual content of A + Structure of B, and \emph{vice versa}.
    \nth{4}: Re-decoration is realized by simply re-mapping to exemplars of different appearance.
    }
    \label{fig:applications}
\end{figure}
In Figure~\ref{fig:applications},
we demonstrate the versatility of our method in several 3D modeling applications with our unified generation framework (more details in the supplementary):
1) \textit{Retargeting:}
The goal is to resize a 3D scene to a target size (typically of a different aspect ratio), while maintaining the local patches in the exemplar.
We simply change the size of the identity mapping field and use it as the initial guess $\synthesisup_0$ without shuffling.
2) \textit{Editing:}
Users can manipulate on a 3D proxy, 
which can be the underlining mapping field or mesh, 
for editing an exemplar or generated scene, such as removal, duplication, and modification.
The manually manipulated proxy is then converted and fed as the initial guess at the coarsest scale for synthesizing the final scene.
3) \textit{Structural analogies:}
Given two scenes A and B, 
we create a scene with
the patch distribution of A, 
but which is structurally aligned
with B. 
This is realized by using the exemplar pyramid of A, and an identity mapping as the initial guess, but
by replacing $\texemplar_0(\synthesisup_0)$ with the transformed features in B, and vice versa.
4) \textit{Re-decoration:}
With the coordinate-based representation,
we can re-decorate the generated ones with ease,
via simply remapping to exemplars of different appearance.

%% file: 6-conclusion.tex
\section{Discussion, Limitations and Future Work}

This work makes an first attempt towards a generic generative model for synthesizing highly realistic general natural scenes from only one exemplar.
Building upon Plenoxels, 
our method can efficiently synthesize diverse and high-quality scenes.
The generated samples particulary inherit photo-realism and view-dependent effects from the example.
Despite success demonstrated, 
we note a few shortcomings.
We can not handle scenes eluding Plenoxels (e.g., transparent fluids, strong reflection),
which is the actual input to our framework.
Particularly, the Plenoxels-based representation is not suitable for large and unbounded scenes, leading to artifacts in the results (more discussion in supplementary).
With voxelized volumetric representations,
we can not perfectly synthesize scenes with tiny thin structures,
and ones with highly semantic or structural information, e.g., human body and modern buildings.
Moreover,
in contrast to \emph{continuous} distributions learned in neural-based methods,
we work on \emph{discrete} patch distributions and thus lack the capability of generating novel patches/pixels.
A future direction is to learn a continuous distribution from a large number of homogeneous samples produced by our method,
with GANs, VQ-VAEs, or diffusion models.
Last,
the view-dependent effects of the results are inherited from the input Plenoxels,
although SH features have already \emph{implicitly} considered the veiw-dependent lighting,
consistent global illumination can not be guaranteed in our results, leading to another future direction.

%% file: 7-supplementary.tex
\section{More Visual Results}
\begin{figure*}[t!]
    \centering
    \includegraphics[width=\hsize]{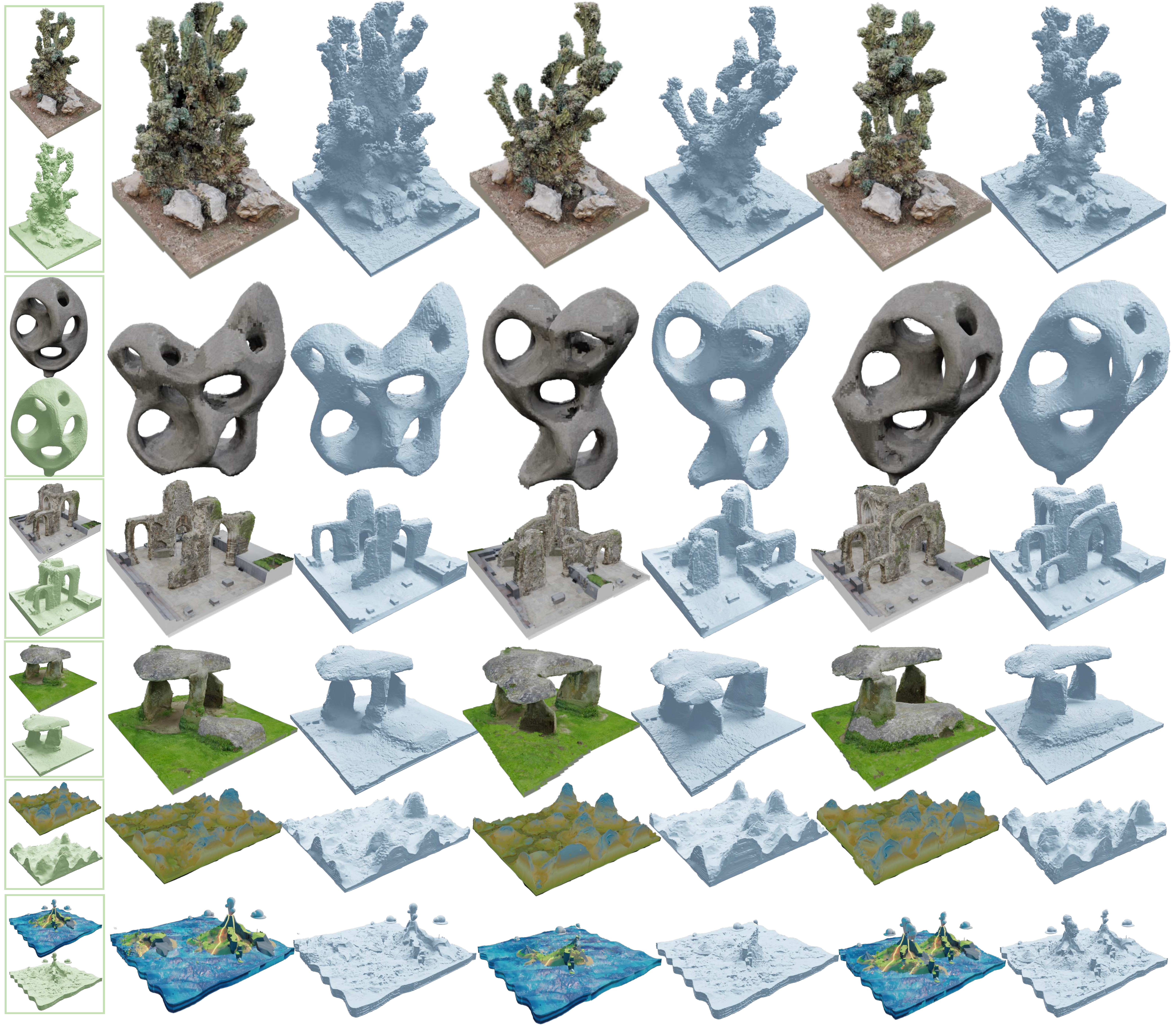}
    \caption{
    Visualization of the high-quality geometries generated by our methods. 
    The original exemplar scenes for generating these results are: (from top to bottom)
    Cactus Cereus~\cite{cactus_cereus}, Stone Sculpture~\cite{stonesculpture}, St Alphage~\cite{alphage}, Spinsters Rock~\cite{spinstersrock}, The Vast Land~\cite{vastland}, Volcano Island Lowpoly~\cite{volcanoislandlowpoly}.
    }
    \label{fig:supplementary_mesh_result}
\end{figure*}

More artistic pieces, that are created with high-quality and diverse general natural scenes generated by our method, are shown in Figure~\ref{fig:supplementary_teaser}.
Moreover,
in Figure~\ref{fig:supplementary_mesh_result},
we also visualize the underlying high-quality and diverse geometry of more generated samples.
Figure~\ref{fig:supplementary_results_0}, ~\ref{fig:supplementary_results_1} presents more samples generated with our method.

In addition to
more dynamic 3D viewing of a large collection of generated samples presented in the supplementary video,
please also see
the anonymous project website
\href{http://wyysf-98.github.io/Sin3DGen}{http://wyysf-98.github.io/Sin3DGen} 
for a more immersive view into our 3D results.

\section{Implementation Details}
All synthesis results presented in this paper share the following default setting, 
unless specified.
\emph{
We will release the code for reproducing the results presented in this paper, 
upon the publication of this work.
}

\paragraph{Random Generation.}

By default,
the synthesized scene $\synthesis$ shares the same bounding box $\bound$ with $\exemplar$ in the random synthesis task.
Each scene is located inside a cuboid, 
of which the aspect ratio varies according to different exemplars.
In Table~\ref{tab:data_configuration}, 
we list the final resolution of $\synthesis_N$ in the pyramidal generation framework for each exemplar scene.
At the $N$-th scale in the multi-scale framework,
the resolution along the maximum dimension of $\synthesis_N$ is set to 121, considering the trade-off between the quality and computational efficiency of the generation.
The resolution along the maximum dimension of the higher-resolution $\exemplar^{high}$ is 512.
The scaling factor between consecutive scales in the pyramid is $r=4/3$, 
and the coarsest resolution is 16.
We use $N=7$ in the pyramid, which results in 8 scales in total.
The patch size at all scales is set to $p=5$.
We set the number of PCA components to 3, the truncate scale of SDF $t = 3 \times w$, 
where $w$ is the voxel size.
The weight of the appearance feature is ${w_a} = 0.5$, 
the completeness trade-off weight $\alpha = 0.01$, 
and the initial noise $\sigma = 0.5$.
At coarser scales ($n < 5$), 
exact NNF is applied $T_e=10$ times, which means the value-based NNF is performed with $T_e-1=9$ times and followed by one mapping-based NNF search;
At the finest scale ($n >= 5$), 
approximate NNF via PatchMatch is performed $T_a=2$ times. 
The “jump flood” radius is 8, and the random search radius is fit to the max resolution of the current exemplar.

\paragraph{Applications.} 
In contrast to the random synthesis task,
the $\sigma$ for noise used in all applications is set to 0.
More specifically:
\begin{itemize}
\item 
1) \textbf{Retargeting:}
The goal is to resize a 3D scene to a target size (typically of a different aspect ratio), while maintaining the local patches in the exemplar. 
We simply set the resolution of $\synthesis_N$ to the target size,
and the resolution of $\synthesis_0, ..., \synthesis_{N-1}$ is adapted accordingly with the default scaling factor $r$.
\item 
2) \textbf{Editing:}
Users can manipulate on a 3D proxy, which can be the underlining mapping field or mesh, for editing an exemplar or generated scene, such as removal, duplication, and modification. 
The manually manipulated proxy is then converted and fed as the initial guess at the coarsest scale for synthesizing the final scene.
As editing the 3D scene requires more meticulous 3D interaction,
we set the resolution at the coarsest scale to a higher value (resolution along the max dimension is 28), and use 6 scales in total.
We perform the exact NNF at the first 3 scales, 
followed by 3 finer scales with the approximate NNF.
\item 
3) \textbf{Structural analogies:}
Given two scenes A and B, 
we create a scene with
the patch distribution of A, 
but which is structurally aligned
with B. 
This is realized by using the exemplar pyramid of A, and an identity mapping as the initial guess, but replacing $\texemplar_0(\synthesisup_0)$ with the transformed features in B, and vice versa.
As the content of the generated scene at the coarsest scale is already specified by $\texemplar_0(\synthesisup_0)$,
the pyramidal generation starts with a higher-resolution scale (51 voxels along the max dimension), finishes the generation with 4 scales in total, and performs exact NNF in the first scale.
\item 
4) \textbf{Re-decoration:}
Trivially,
we do not need to re-synthesize the scene in the re-decoration application.
Given an already generated scene,
a novel scene can be obtained by simply remapping the coordinate-based synthesis result to an exemplar of different appearance.
\end{itemize}

\input{table/supplementary_data_configuration.tex}

\input{table/supplementary_data_configuration_application.tex}

\section{Datasets}
We collected a rich variety of 3D scene models to examine the performance of our method on random scene generation, 
ranging from rocks to plants, sculptures, landscapes, terrains, artistic scenes, etc.
For each 3D scene model, 
we render $200$ images at the resolution 1024 × 1024, with cameras distributed on a sphere in Blender~\cite{blender}.
Then the Plenoxels-based exemplar pyramid is obtained via coarse-to-fine training on these images.
Notably,
in Figure~\ref{fig:realworld_data},
we also demonstrate our method on real-images collected from a real-world scenic site.
To this end, 
we collect $300$ images with the resolution 1280 x 720 from Google Earth Studio~\cite{google_earth_studio}, 
where we can manually specify cameras for simulating a drone programmed to fly over a scenic spot,
for training the Plenoxels-based exemplar.
Specifically, we move the camera in a spiral motion and gradually elevate the camera from a high
altitude to a low altitude. 
Then, we use COLMAP~\cite{colmap1, colmap2} to estimate the camera parameters. 
More details can be found in the video. 
Figure~\ref{fig:supplementary_exemplar}
presents the visuals of all exemplars used in this paper.

\section{More Analysis}
In general,
our method is robust to varying hyper-parameters to some extent.
We shall show the effects of using different parameters in the following.

\begin{figure}[h!]
    \centering
    \includegraphics[width=\linewidth]{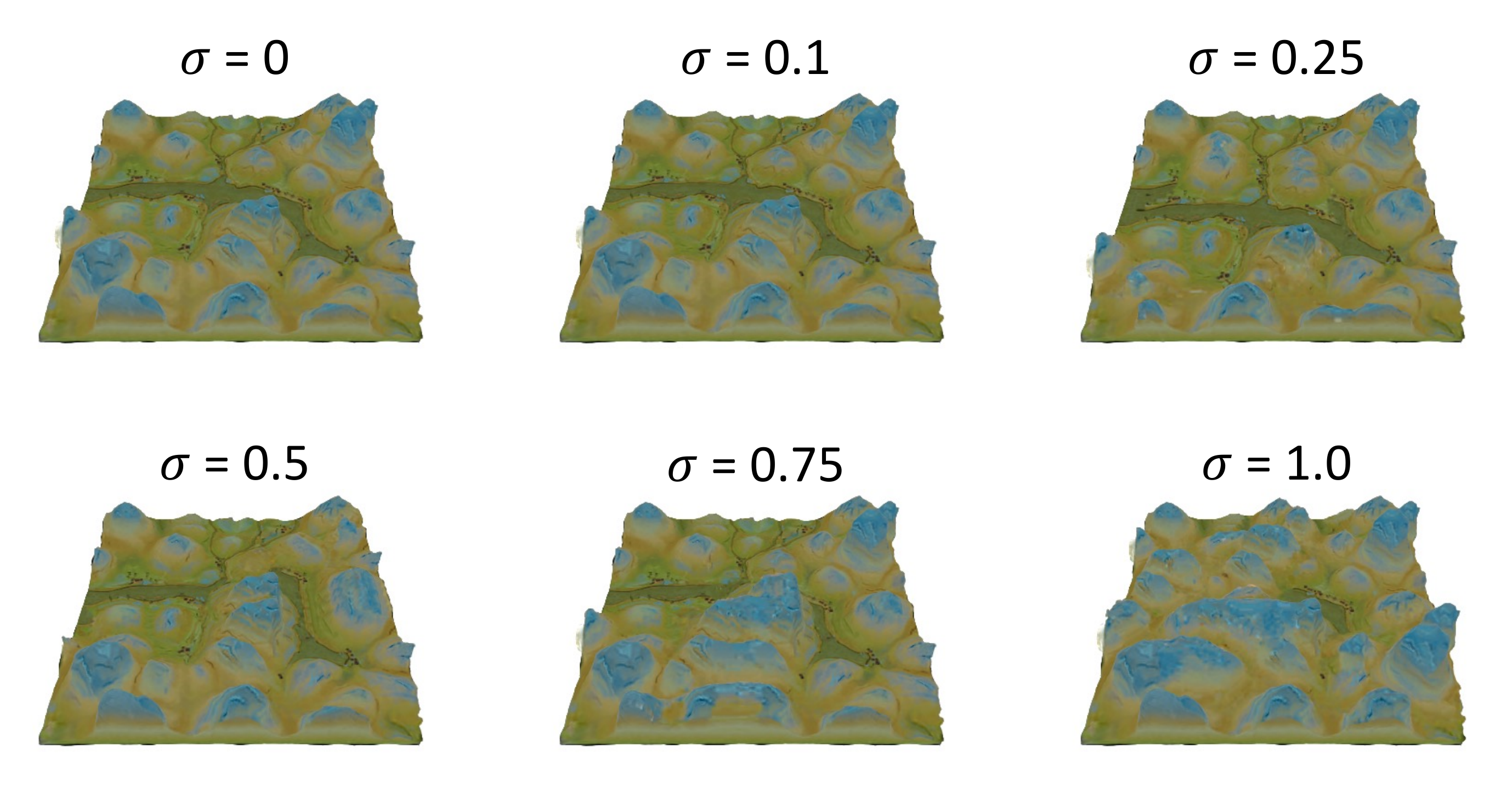}
    \caption{
    Effects of different initial noise.
    Intuitively,
    noise with smaller $\sigma$ values leads to more similar scenes to the exemplar,
    while larger values result in more diverse ones.
    }
    \label{fig:effects_noise}
\end{figure}

\paragraph{Effects of Different Noise $z_0$.} 
In Figure~\ref{fig:effects_noise},
we show the results obtained by different initial guesses, i.e.,\ the identity mapping shuffled with different noises, at the coarsest level.

\begin{figure}[t!]
    \centering
    \includegraphics[width=\linewidth]{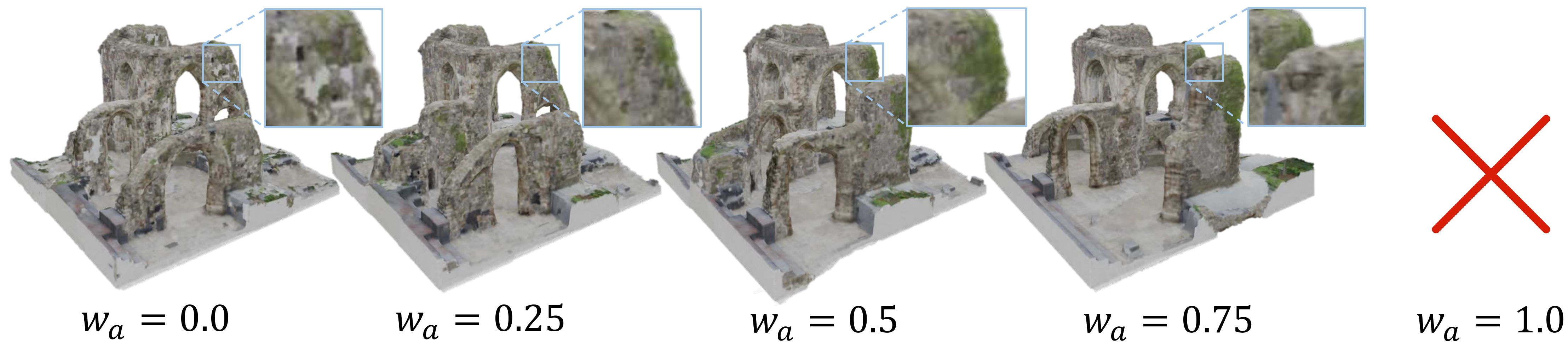}
    \caption{
    Effects of varying $w_a$.
    While $w_a$ around 0.5 produces relatively stable results, extreme values introduce visual inconsistency and artifacts (see left $w_a$), or even fail (see right $w_a=1.0$).
    }
    \label{fig:w_a_effects}
\end{figure}

\paragraph{Effects of Different $w_a$.}
Empirically, the trade-off parameter $w_a$ for balancing the appearance and geometry feature is set to 0.5 by default.
While we have shown this setting yields robust and high-quality generation, 
we shall also demonstrate the effects of varying weights.
Generally,
in Figure~\ref{fig:w_a_effects}, 
we can see that our method is robust to varying $w_a$ to some extent.
However,
setting $w_a$ to an extremely small value, which pays much attention to the geometry feature, will lead to inconsistent appearance and artifacts. 
On the other end,
if we put all emphasis on the geometry feature, the synthesis fails and produces empty scenes.

\begin{figure}[t!]
    \centering
    \includegraphics[width=\linewidth]{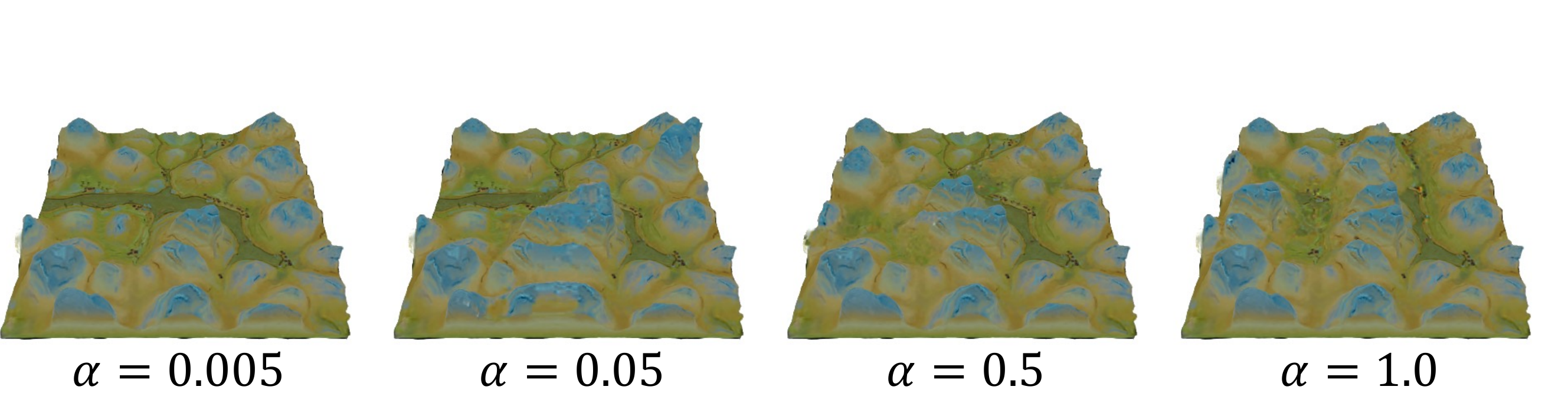}
    \caption{
    Effects of varying $\alpha$.
    $\alpha$ serves as a coarse control knob for visual completeness.
    The river becomes shorter, suggesting lower visual completeness, as $\alpha$ increases.
    }
    \label{fig:alpha_effects}
\end{figure}

\paragraph{Effects of Different $\alpha$.}
Figure~\ref{fig:alpha_effects} presents the effects of varying $\alpha$ for different levels of visual completeness.
Nevertheless, 
we also found the degrees of such control may not be always perfectly explicit, which we also observed with~\cite{drop}.

\begin{figure}[t!]
    \centering
    \includegraphics[width=\linewidth]{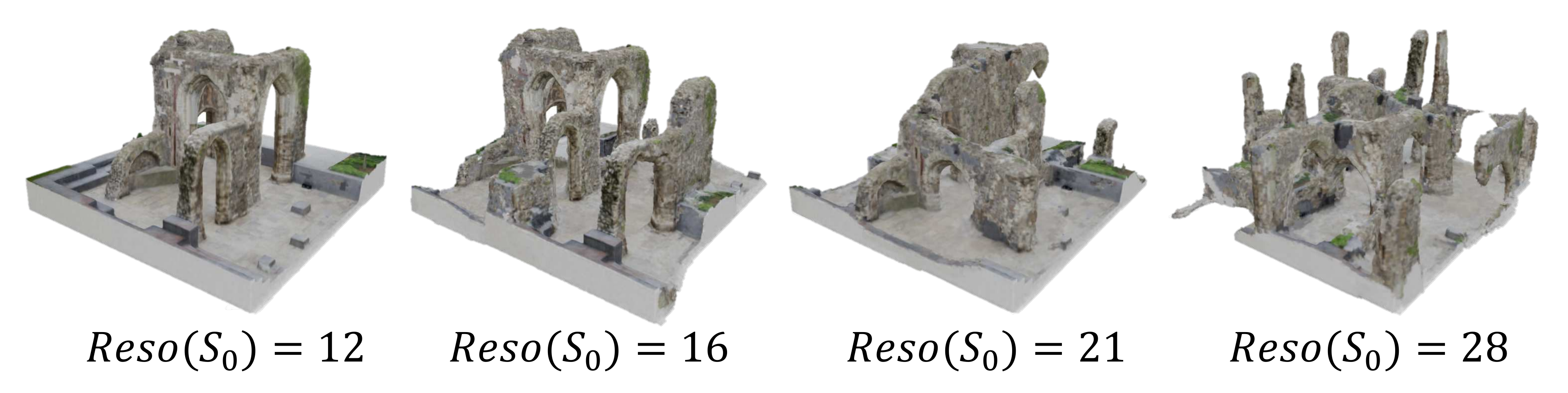}
    \caption{
    Effects of different resolutions of $\synthesis_0$. Lower-resolution $\synthesis_0$ (larger receptive field at coarse scale) results in less structural diversity, 
    producing almost identical to the exemplar. 
    On the contrary, 
    with smaller receptive fields at the coarsest scale, the global arrangement can not be well preserved (see messy structures on the right).
    }
    \label{fig:s_0_effects}
\end{figure}

\paragraph{Effects of Different Resolution for $\synthesis_0$ at the Coarsest Scale.}
Given a fixed patch size, which is $p=5$ in our work,
a larger resolution at the coarsest scale suggests a smaller effective receptive field (the same concept as in convolutional neural networks)  
and less-considered global layouts at the coarser scales, 
and vice versa.
In our work, 
we by default use the setting where the patch at the coarsest scale captures 1/3 of the content in the exemplar, 
balancing the local diversity and global layout. 
In Figure~\ref{fig:s_0_effects}, 
we also show the impact of varying resolutions at the coarsest scale, that capture contents of different sizes in the generation.

\begin{figure}[t!]
    \centering
    \includegraphics[width=\linewidth]{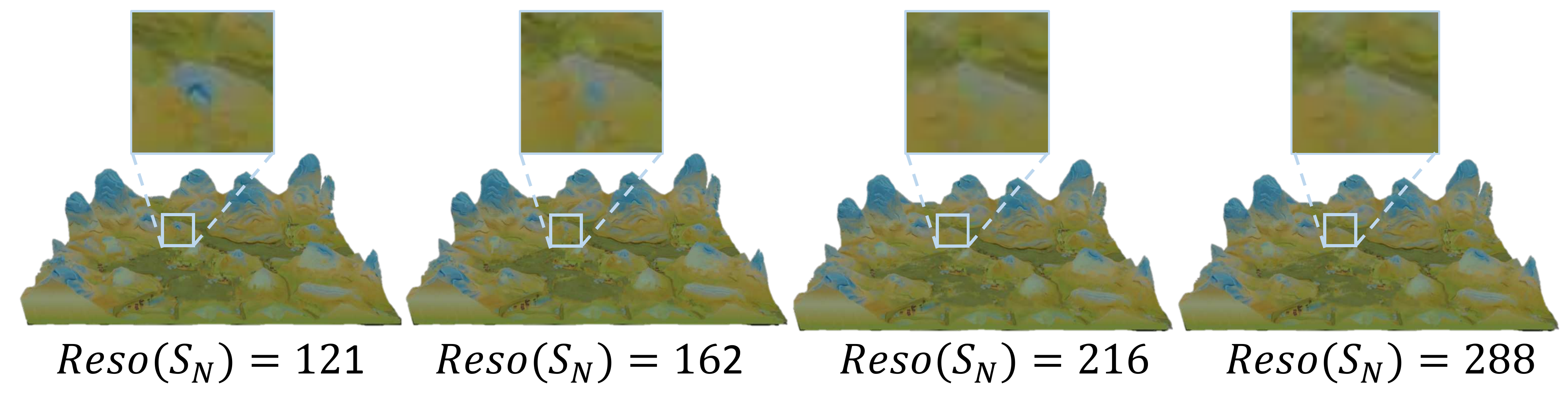}
    \caption{
    Effects of different resolutions of $\synthesis_N$. The overall visual synthesis is stable at a resolution of 121. Some minor aliasing can be found in lower resolution (left), with a larger scale for synthesis, more details can be obtained.
    }
    \label{fig:s_N_effects}
\end{figure}

\paragraph{Effects of Different Resolution of $\synthesis_N$ at the Finest Scale.}
The synthesis at finer scales only considers visual coherence and adds local details. 
We have shown that
synthesizing with a maximum resolution 121 by default in the pyramid
is sufficient in most cases for the trade-off between quality and efficiency. 
Moreover,
in Figure~\ref{fig:s_N_effects}, 
we show that using higher resolutions for $\synthesis_N$ only leads to negligible visual gains.
Besides, 
we also observed that
as we use approximate NNF at finer scales, 
the inaccurate NNF search may introduce some wrong patches and lead to performance degradation in the generation.

\begin{figure}[t!]
    \centering
    \includegraphics[width=\linewidth]{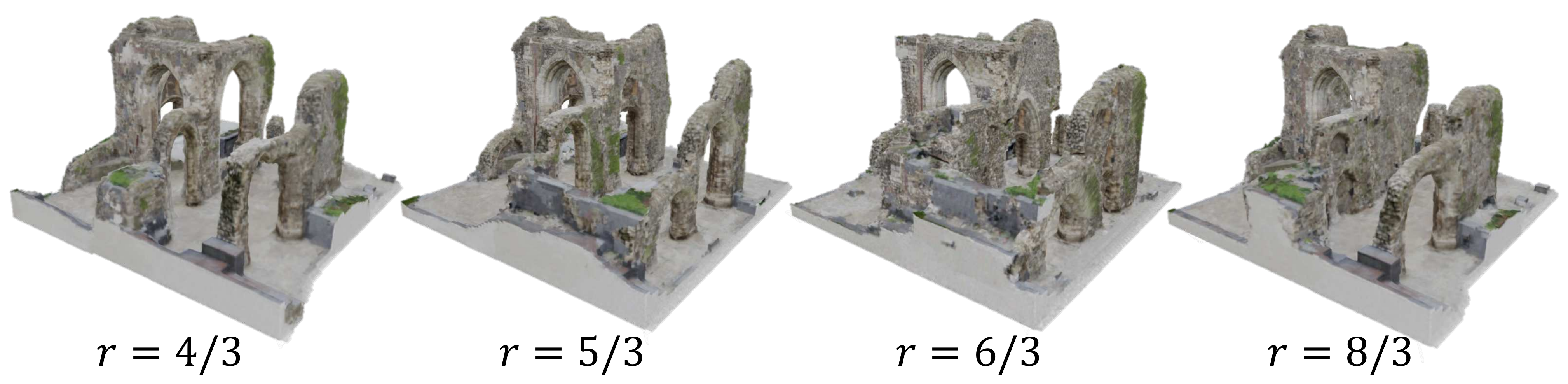}
    \caption{
    Effects of different downscale ratios $r$. As $r$ increasing, fine structures, e.g., the arch doors, gradually disappears, producing bulky structures and less diversity in each instance.
    }
    \label{fig:r_effects}
\end{figure}

\paragraph{Effects of Different Downscale Ratio $r$.}
The downscale ratio $r$ used for building the pyramid affects the transition between scales. 
As the ratio increases, the transition of the generation between scales becomes more inconsistent and unstable due to large gaps between consecutive scales,
leading to the loss of fine structures and less diversity as shown in Figure~\ref{fig:r_effects}.

\begin{figure}[t!]
    \centering
    \includegraphics[width=\linewidth]{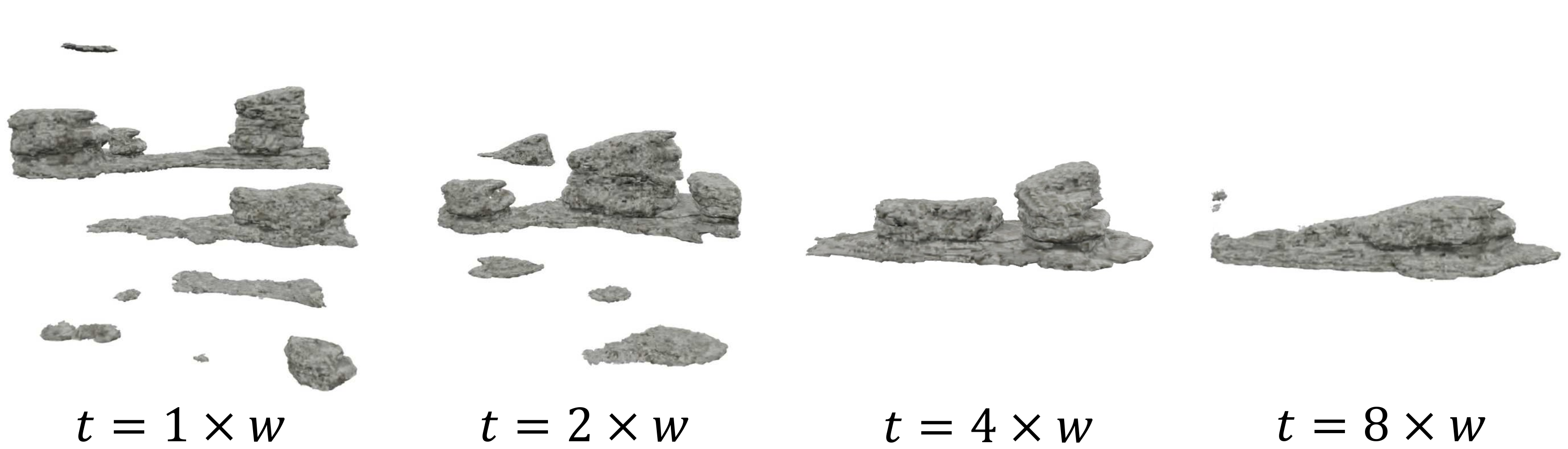}
    \caption{
    Effects of different truncate scales $t$, where $w$ is the current voxel size at each scale.
    A small $t$ will lead to separated fragments due to the loss of geometry information.
    On the contrary, a very larger truncate scale will blur the geometric feature and only synthesize instances with a coarse shape.
    }
    \label{fig:truncated_scale_effects}
\end{figure}

\paragraph{Effects of the Truncated Scale $t$.}
The truncated scale $t$ controls
the range of geometric features we keep for patching matching.
Smaller truncate scales only consider information near to surface and degrade to the occupancy field, which may produce many tiny pieces and incomplete instances,
while larger values lead to blurry results.
Figure~\ref{fig:truncated_scale_effects} presents the visual results.

\paragraph{Only Exact or Approximate NNF-3D.}
The mix use of exact NNF and approximate NNF in our framework has shown the efficacy and efficiency in 3D generation.
Using only exact NNF would quickly lead to prohibitive computational cost
and prevent us from synthesizing high-resolution results.
See Table~\ref{tab:computational_overhead} for the detailed computation overhead.
On the other end,
only using approximate NNF all the time
will harm the generation, producing distorted results,
as the approximate NNF is inaccurate.
In Figure~\ref{fig:truncated_scale_effects}, we show the visual results when only using approximate NNF-3D.

\begin{figure}[t!]
    \centering
    \includegraphics[width=\linewidth]{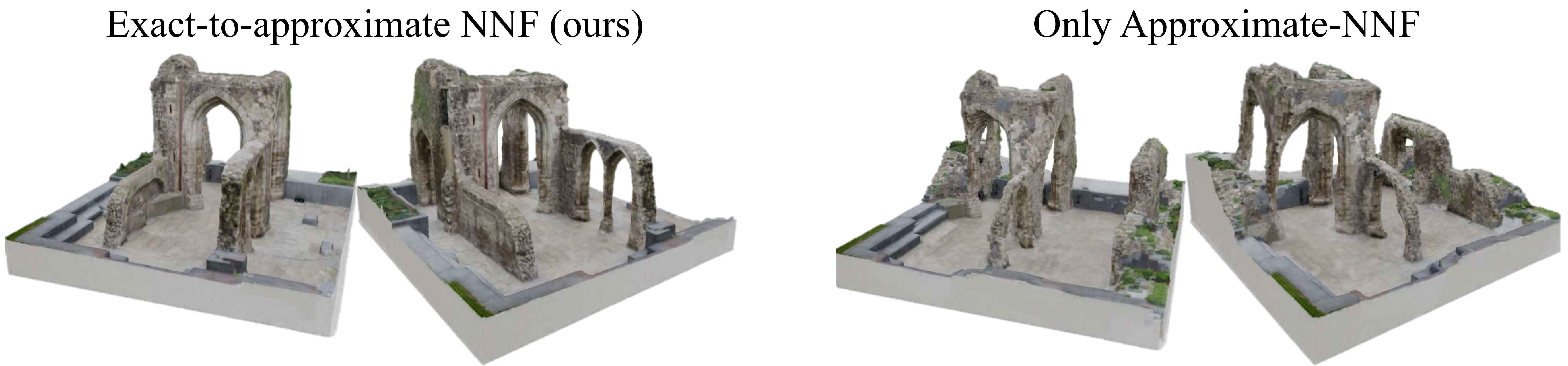}
    \caption{
    Approximate NNF-only generation. 
    See the distorted arches occurred in the results on the right.
    }
    \label{fig:only_approximate}
\end{figure}

\section{More Experiments}
\begin{figure}[t!]
    \centering
    \includegraphics[width=\linewidth]{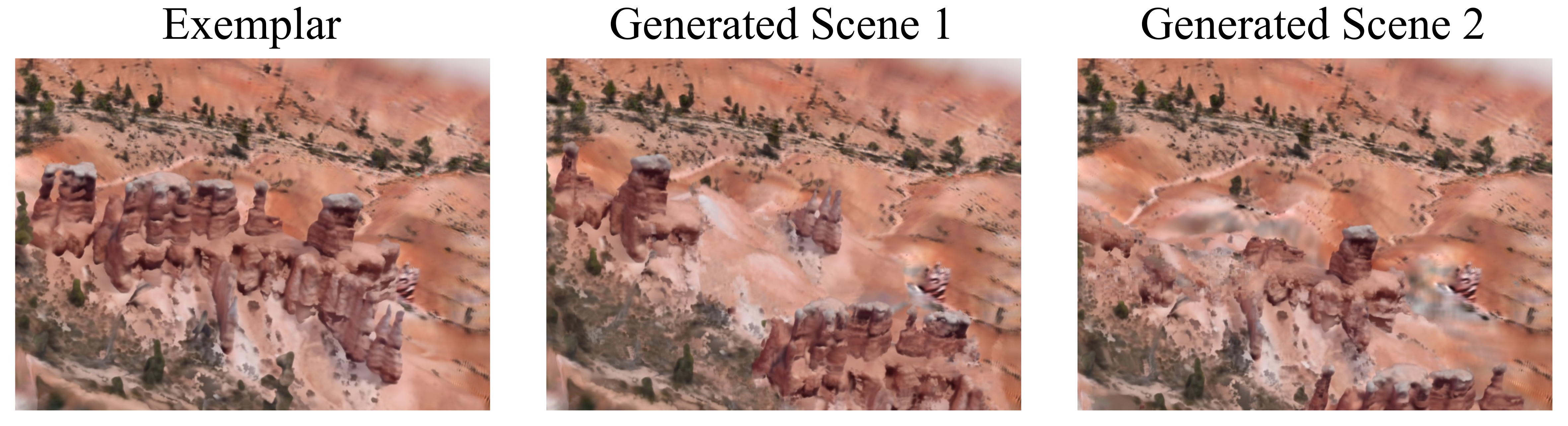}
    \caption{
    Samples generated with images collected from a real-world scenic site \--- Bryce canyon~\href{https://www.google.com/help/terms_maps/}{\textcopyright2022 Google}. Notably, we only synthesize the region of interest (i.e., the odd rocks) and the background is disentangled out by modeling with an independent implicit neural network.
    }
    \label{fig:realworld_data}
\end{figure}

\paragraph{Working with Unbounded Scenes.}
Benefiting from using Plenoxels, 
which trains on 2D images,
for representing the input scene,
our method can also work on images collected from a \emph{real-world unbounded} scene. 
To this end, 
we use COLMAP~\cite{colmap1, colmap2} to estimate the camera parameters,
and model the background using an implicit neural network, similar to NeRF++~\cite{nerfpp}.
Figure~\ref{fig:realworld_data} presents the results, more visual details can be found in the video. 
Note that,
existing NeRF-based models often struggle in handling "unbounded" real-world scenes, and disentangling the foreground and background. 
Nevertheless,
some works~\cite{barron2022mipnerf360, lumaai} attempt to tackle these problems, 
showing promising results.
We believe these methods can help boost the performance of our method on more real-world scenes, which, however, is not in the scope of this work and stimulates future research.

\paragraph{Computational Overhead.}
In Table~\ref{tab:computational_overhead},
we reported the detailed time and memory usage for the exact-only NNF and approximate-only NNF.
As aforementioned,
using either exact-only or approximate-only NNF would not be satisfying,
and our exact-to-approximate scheme is the key to enable synthesizing high-quality results with limited computational resources.
\input{table/supplementary_computational_overhead.tex}

\begin{figure}[t!]
    \centering
    \includegraphics[width=\linewidth]{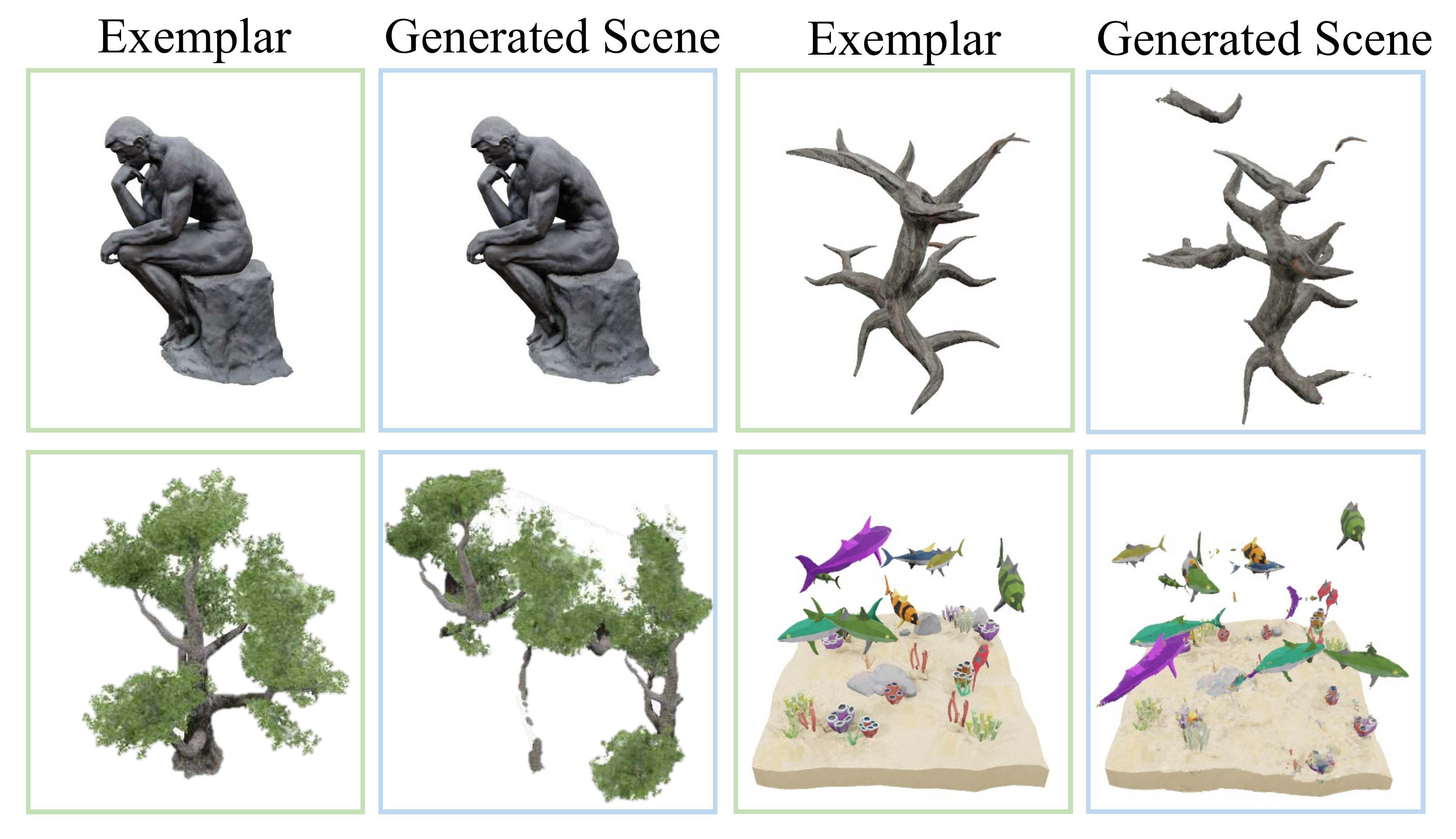}
    \caption{
    Failure cases. 
    An exemplar scene,
    that does not have sufficient diverse patch candidates, would result in identical generation results to the input (See top left sculpture~\cite{derdenker}).
    Our method failed on scenes with tiny thin structures, such as branches~\cite{olddeadtree} and trees~\cite{morerealistictrees}.
    Scenes with highly semantic or structural information can not be correctly handled (See broken fishes on the bottom right~\cite{diving, lowpolyfish}).
    }
    \label{fig:failure_cases}
\end{figure}

\paragraph{Failure Cases.}
As mentioned in the paper, 
our method favors scenes with complex structures
and textures for matching the internal distribution, lacking sufficient diverse patch candidates will lead to mode collapse issues. 
Besides, 
with voxelized volumetric representations,
our method can not perfectly synthesize scenes with tiny thin structures.
Moreover,
our method operates on the patch level, 
so we can not guarantee
that highly semantic or structural features in the exemplar can be preserved intactly in the synthesized results.
Figure~\ref{fig:failure_cases} shows some failure cases when working with our method.

\section{Evaluation}

\subsection{Baselines}
\paragraph{GRAF \cite{graf}} 
We use the official implementation,
and replace the camera poses with ones in our work. 
We follow the default setting for training, 
one model for each exemplar scene is trained with renderings of resolution $512 ^{2}$ for $7200k$ samples, 
which takes about 3 days in a single V100 GPU.
The final visuals are rendered at the resolution $512^2$.

\paragraph{StyleNeRF \cite{stylenerf}} 
We use the official release of StyleNeRF. 
Same as GRAF, we replace the camera extrinsic and intrinsic parameters with the real distribution and set the background to white. 
All models are trained in the resolution of $512 ^{2}$ following the default setting by going through $6000k$ images for about 3 days using 4 V100 GPUs. 

\paragraph{GPNN-3D} 
We naively extend the GPNN~\cite{drop} for working on Plenoxels-based exemplar scenes. 
The density value and SH values are normalized to fit $[-1, 1]$, 
and we follow all parameters as described in the original paper. 
The maximum resolution reached by GPNN-3D is only 38 due to computational efficiency issues. 

\subsection{Camera Pose Sampling}
To quantitatively evaluate the synthesized scenes from 2D projections, 
we uniformly sample $K = 50$ camera poses on the upper hemisphere with radius $R = 2.5$,
and use elevation angles range from $0^{\circ}$ to $90^{\circ}$. 
The focal length of the camera is set to $512$ times the pixel size, equivalently FoV $\approx 39.6^{\circ}$, for all exemplar scenes.

\subsection{Metrics}
For each method,
we produce $50$ generated scenes on each of the evaluated exemplars,
render 50 multi-view images and extract the 3D geometric surface points of the exemplar and each in the generated,
and then rate the performance using a combination of several common metrics in both 2D and 3D generation:
  \paragraph{Visual Quality} measures how well the model captures the internal statistics of the input exemplar from the 2D perspective, by simply computing the averaged SIFID~\cite{singan} over multiple views of a generated scene.
  Concretely,
  for each image rendered from a generated scene, we compute the single image SIFID of this image against the image rendered at the associate viewpoint in the exemplar scene.
  Then the SIFID-MV for a generated scene is the average over the multiple views.
  We finally report the mean SIFID-MV averaged over multiple generated scenes.

  \paragraph{Visual Diversity} of the set of generated scenes is measured via extending the image diversity score as in~\cite{singan} to multi-view images of a scene.
  First,
  under each view,
  we calculated the standard deviation (std) of the intensity values of each pixel over 50 images rendered from 50 generated scenes, averaged it over all pixels, and normalized by the std of the intensity values of the image rendered from the exemplar.
  Then, we report the average of std values obtained at 50 views as the Visual Diversity of a set of generated scenes.
  
  \paragraph{Geometry Quality} of a generated scene is measured as the Minimal Matching Distance~\cite{mmsc} (multiplied by $10^2$) between the set of generated patches and exemplar patches (represented as point clouds sampled on the surface).
  As mentioned in the paper,
  Plenoxels often produce invisible noise, so we only pick point cloud patches on the surface.
  Specifically,
  for a scene represented by a discrete volume of resolution $256^3$,
  we extract mesh using Marching Cubes~\cite{marchingcubes}, and evenly sample 102400 points from the mesh surface.
  To extract patches, we randomly pick 1000 points center, then combined them with the nearest 1024 points via k-NN search.
  Then the geometry quality of a generated scene is calculated as the MMD between the set of patches in a generated scene and the set of patches in the exemplar scene.
  We then report the averaged geometry quality score over 50 generated scenes.
  
  \paragraph{Geometry Diversity} of generated scenes is measured by summing up all the differences among the $50$ generated scenes, i.e.,\ Total Mutual Difference as in~\cite{mmsc}. 
  Specifically,
  we evenly sample 10240 points on the surface of each generated mesh to obtain a point cloud, forming a set of 50 scene point clouds. Empty scenes are deprecated to calculate the geometry diversity.
  Then the Geometry Diversity of 50 generated scenes is reported as the TMD calculated on this set of point clouds.

%% file: table/supplementary_data_configuration.tex
\begin{table}[t!]
\caption{
Resolution configuration for figures in the \textbf{main paper}.
}
\footnotesize
\centering

\begin{tabular}{lll} 
\toprule	
Figure & Data & Resolution of $S_N$ \\ 
\midrule	
Fig. 1 & Cactus Cereus~\cite{cactus_cereus}& $92\times108\times121$ \\ 
\midrule
\begin{tabular}[c]{@{}l@{}}Fig. 2 \&\\Fig. 5\end{tabular} & Green Island~\cite{greenisland} & $121\times121\times47$ \\ 
\midrule
\begin{tabular}[c]{@{}l@{}}Fig. 3 \&\\Fig. 5\end{tabular} & St Alphage~\cite{alphage} & $121\times121\times92$ \\ 
\midrule
\multirow{9}{*}{Fig. 5} & Calda House~\cite{caldahouse} & $121\times121\times71$ \\
 & Callanish~\cite{callanish} & $121\times121\times47$ \\
 & Stone Arch~\cite{longarchstone} & $121\times51\times71$ \\
 & Desert Lowpoly~\cite{lowpolydesert} & $121\times121\times92$ \\
 & Meteora~\href{https://www.google.com/help/terms_maps/}{\textcopyright2022 Google} & $121\times121\times47$ \\
 & Spinsters Rock~\cite{spinstersrock} & $121\times121\times84$ \\
 & Stone Sculpture~\cite{stonesculpture} & $108\times84\times121$ \\
 & Volcano Island Lowpoly~\cite{volcanoislandlowpoly} & $121\times121\times71$ \\
 & The Vast Land~\cite{vastland} & $121\times121\times47$ \\ 
\midrule
Fig. 7 & Mountain with Lakes~\cite{grassymountainswithlakes} & $121\times121\times72$ \\ 
\midrule
\multirow{2}{*}{Fig. 10} 
 & Autumn Camping~\cite{autumncamping} & $121\times99\times72$ \\
 & Winter Camping~\cite{wintercamping} & $121\times99\times72$ \\
\midrule
\multirow{4}{*}{\begin{tabular}[c]{@{}l@{}}Fig. 5 \&\\Fig. 10\end{tabular}} 
 & Devil's Tower~\href{https://www.google.com/help/terms_maps/}{\textcopyright2022 Google} & $121\times121\times63$ \\ 
 & Cactus Saguaro~\cite{cactussaguaro} & $71\times72\times121$ \\
 & Camping Lowpoly~\cite{campinglowpoly} & $121\times99\times72$ \\
 & Mountain~\cite{blackmountain} & $121\times121\times84$ \\
 & Stylized Cactus~\cite{stylizedcactus} & $121\times121\times121$ \\
\bottomrule	

\end{tabular}

\label{tab:data_configuration}
\end{table}%

%% file: table/supplementary_data_configuration_application.tex
\begin{table}[t!]
\caption{
Resolution configuration for \emph{high-resolution synthesis} and \emph{the retargeting application} in the \textbf{main paper}.
}
\footnotesize
\centering

\begin{tabular}{lll} 
\toprule	
Figure & Exemplar & Resolution of $S_N$ \\ 
\midrule	
Fig. 6 & The Vast Land~\cite{vastland} & $288\times288\times112$ \\ 
\midrule
\multirow{4}{*}{Fig. 10} 
 & Cactus Cereus~\cite{cactus_cereus} & $92\times108\times152$ \\
 & Stone Arch~\cite{longarchstone} & $242\times51\times71$ \\
 & Tiny Castle~\cite{tinycastle} & $121\times63\times237$ \\
 & Train Wagon~\cite{trainwagon} & $47\times182\times47$ \\ 
\bottomrule	

\end{tabular}

\label{tab:data_configuration_application}
\end{table}%

%% file: table/supplementary_computational_overhead.tex
\begin{table}[t!]
\caption{
Computational overhead. We report the time and memory consumption in each NNF iteration for exact-only and approximate-only NNF.
}
\footnotesize
\setlength{\tabcolsep}{2.5pt}
\centering

\begin{tabular}{@{}l|cc|cc@{}}
\toprule
\multirow{2}{*}{Resolution} & \multicolumn{2}{c|}{Only Exact NNF} & \multicolumn{2}{c}{Only Approximate NNF} \\ \cmidrule(l){2-5} 
 & Time (s) & Memory (GB) & Time (s) & Memory (GB) \\
 \midrule
$16\times16\times5$ & 0.01 & 0.07 & 0.55 & 0.07 \\
$21\times21\times7$ & 0.01 & 0.12 & 0.59 & 0.14 \\
$28\times28\times10$ & 0.02 & 0.68 & 0.69 & 0.21 \\
$38\times38\times14$ & 0.08 & 1.48 & 0.98 & 0.43 \\
$51\times51\times19$ & 0.57 & 3.66 & 1.69 & 0.92 \\
$68\times68\times26$ & 4.06 & 9.74 & 3.62 & 2.02 \\
$91\times91\times35$ & 27.47 & 25.17 & 8.47 & 4.77 \\
$121\times121\times47$ & N/A & N/A & 20.07 & 11.28 \\
$162\times162\times63$ & N/A & N/A & 48.91 & 17.05 \\
$216\times216\times84$ & N/A & N/A & 116.09 & 21.95 \\
$288\times288\times112$ & N/A & N/A & 278.53 & 23.27 \\
$384\times384\times150$ & N/A & N/A & 662.23 & 26.60 \\ 
\bottomrule
\end{tabular}

\label{tab:computational_overhead}
\end{table}%